\documentclass[twocolumn, usenatbib]{mnras}
\pdfoutput=1 
\usepackage{amsmath,amstext}
\usepackage{appendix}
\usepackage[T1]{fontenc}
\usepackage{ae,aecompl}
\usepackage[figure,figure*]{hypcap}
\usepackage[graphicx]{realboxes}
\usepackage{amssymb}	
\usepackage[caption=false]{subfig}
\usepackage{caption}
\usepackage{rotating}
\usepackage{threeparttable}

\usepackage{tikz}

\newcommand{\fluxunits}{$10^{-17}$ erg/s/cm$^2$/\AA}

\newcommand{\noop}[1]{}
\definecolor{red75}{rgb}{0.75, 0.0, 0.0}
\definecolor{green75}{rgb}{0, 0.75, 0.0}

\usepackage{soul,xcolor}
\setstcolor{red}

\makeatletter
\renewcommand*{\@fnsymbol}[1]{\ifcase#1\or ^\dagger\or ^{\star} \else\@ctrerr\fi}
\makeatother
\title[MaNGA: Strong Galaxy Lens Candidates]{SDSS-IV MaNGA: A Catalogue of Spectroscopically Detected Strong Galaxy-Galaxy Lens Candidates}

\author[Talbot and Brownstein et al.]{
\parbox{\textwidth}{Michael S.\,Talbot,$^{1}$\thanks{E-mail: \texttt{\href{mailto:michaeltalbot@astro.utah.edu}{michaeltalbot@astro.utah.edu}}}
Joel R.\,Brownstein,$^{1}$\thanks{E-mail: \texttt{\href{mailto:joelbrownstein@physics.utah.edu}{joelbrownstein@physics.utah.edu}}}
Justus Neumann,$^{2}$
Daniel Thomas,$^{2}$
Claudia Maraston,$^{2}$
Niv Drory$^{3}$
\\}
\\
$^{1}$Department of Physics and Astronomy, University of Utah, 115 S. 1400 E., Salt Lake City, UT 84112, USA \\
$^{2}$ Institute of Cosmology \& Gravitation, University of Portsmouth, Dennis Sciama Building, Portsmouth, PO1 3FX, UK \\
$^{3}$ McDonald Observatory, The University of Texas at Austin, 1 University Station, Austin, TX 78712, USA \\
}

\date{Accepted 20 June 2022. Received 13 August 2021.}
\pubyear{2022}
\begin{document}
\maketitle
\begin{abstract}

We spectroscopically detected candidate emission-lines of 8 likely, 17 probable, and 69 possible strong galaxy-galaxy gravitational lens candidates found within the spectra of $\approx10,000$ galaxy targets contained within the completed Mapping of Nearby Galaxies at Apache Point Observatory (MaNGA) survey. This search is based upon the methodology of the Spectroscopic Identification of Lensing Objects (SILO) project, which extends the spectroscopic detection methods of the BOSS Emission-Line Lensing Survey (BELLS) and the Sloan Lens ACS Survey (SLACS). We scanned the co-added residuals that we constructed from stacks of foreground subtracted row-stacked-spectra (RSS) so a sigma-clipping method can be used to reject cosmic-rays and other forms of transients that impact only a small fraction of the combined exposures. We also constructed narrow-band images from the signal-to-noise of the co-added residuals to observe signs of lensed source images. We also use several methods to compute the probable strong lensing regime for each candidate lens to determine which candidate background galaxies may reside sufficiently near the galaxy centre for strong lensing to occur. We present the spectroscopic redshifts within a value-added catalogue (VAC) for data release 17 (DR17) of SDSS-IV. We also present the lens candidates, spectroscopic data, and narrow-band images within a VAC for DR17. High resolution follow-up imaging of these lens candidates are expected to yield a sample of confirmed grade-A lenses with sufficient angular size to probe possible discrepancies between the mass derived from a best-fitting lens model, and the dynamical mass derived from the observed stellar velocities.

\end{abstract}

\section{Introduction}\label{ch4intro}

Strong galaxy-galaxy gravitational lenses sufficiently alter the path of light from a source galaxy, which can be spectroscopically observed at significantly higher redshifts than the target galaxy, to cause multiple images of the source to converge along the line-of-sight (LOS). The strength of the deflection depends on the enclosed mass. Since the image resolution of space and some ground telescopes can be an order of a tenth to a hundredth of an arcsecond (\arcsec), the arcsecond-scale geometry of source images can be resolved and modelled to project the enclosed lens mass at cosmological distances.

The orbital velocities of stars depend on the gravitational potential, which enables a second direct measurement of the total mass enclosed within a central spherical region of the galaxy. However, only a single galaxy-scale measurement of the overall kinematics is typically obtained for galaxies at cosmological distances since the galaxy's angular size is approximately the same as the angular radius of the aperture used to collect the spectra. Without dynamic modelling of spatially resolved line-of-sight (LOS) velocity and stellar dispersion features, the magnitude of the rotational velocity component along the line of sight that causes a relativistic Doppler shift in the spectra, and the magnitude of the random stellar motions that causes a relativistic Doppler broadening of the spectra, cannot be isolated from the unknown orientation and anisotropy of the stellar velocity field that also impacts the shift and broadening in the spectra. This issue is known as the mass-anisotropy degeneracy~\citep{1993MNRAS.265..213G}, which limits dynamic mass measurements of galaxies at cosmological distances to be approximate at best.

The enclosed lensing mass can be used to constrain the magnitude of the stellar velocity model, so the dependence of the modelled stellar velocity field on radius can be adjusted until the spectra broadening predicted by the model matches observations. This joint lens-dynamic fit enables the power-law slope of the radial-dependent density profile of galaxies to be constrained at cosmological distances~\citep{2006ApJ...640..662T, 2008ApJ...684..248B, 2009MNRAS.399...21B, 2010ApJ...724..511A}. The constraints on the power-law slope of the density profile can then be compared to or reproduced by galaxy simulations~\citep{2016MNRAS.456..739X, 2018MNRAS.475.2403L, 2018MNRAS.479.4108M, 2020MNRAS.496.1718E, 2021MNRAS.504.3455M} to balance the competing effect of dark matter, galaxy merging, baryon interactions, and other processes that either tend to broaden or steepen the slope of the profile.

Samples of lenses have also been used to statistically infer the evolution in the mass profile across distance or galaxy radius~\citep{2006ApJ...649..599K, 2011ApJ...727...96R, 2012ApJ...757...82B, 2018MNRAS.480..431L}. Constraints on the lens mass profile are also vital to resolve the tension in the measured expansion rate of the Universe ($H_0$) between the late~\citep{2009ApJ...695..287R, 2018ApJ...857...67H, 2018AAS...23231902P, 2019ApJ...882...34F, 2019ApJ...876...85R} and early Universe~\citep{2018MNRAS.480.3879A, 2020A&A...641A...6P} measurements~\citep{2019NatAs...3..891V}. In particular, $H_0$ can be measured from its impact on the time-delay observed between source images once the fraction of the time-delay caused by the different travel paths are measured~\citep{2015A&A...580A..38R, 2017MNRAS.465.4895W, 2019MNRAS.484.4726B, 2019ApJ...876..107P, 2019MNRAS.490.1743C, 2020ApJ...895L..29L, 2020A&A...639A.101M, 2020MNRAS.496.3270M, 2020MNRAS.498.1440R, 2020MNRAS.494.6072S, 2020ApJ...897..127W, 2020MNRAS.498.1420W, 2020MNRAS.497L..56Y}.

However,~\cite{2018ApJ...855...64L} discovered that the lens mass is statistically 20.7\% higher than the dynamic mass for lenses between the redshift range of $0.1 < z < 0.7$, which implies a modelling bias exists that might impact the quality of the lens, dynamic, or joint lens-dynamic mass measurements. The most suspected cause.

Test here.

of the lens-dynamic mass discrepancy is extra mass either within the lens environment~\citep{1980ApJ...236..351D, 2004ApJ...612..660K, 2005tyad.confE..12D, 2009ApJ...690..670T, 2011ApJ...726...84W, 2014MNRAS.439.2432J, 2018ApJ...867..107W} or along the LOS~\citep{2007arXiv0706.3098G, 2014MNRAS.439.2432J, 2018ApJ...855...64L} has not been modelled as part of the enclosed mass responsible for the deflection of the source light. The LOS mass cannot be isolated by fitting a lens model to observations since a 'sheet' of mass can be traded with the 'mass' scale of the lens model without changing the size or shape of the source images predicted by the model. This issue is known as the mass-sheet transform (MST) or mass-sheet degeneracy (MSD)~\citep{1985ApJ...289L...1F}. Thus LOS mass may contaminate measurements of the lens mass profile~\citep{2007arXiv0706.3098G, 2007ApJ...660L..31M}. The uncertainty in the measurement of $H_0$ is inflated by uncertainties in the impact of the LOS mass and lens profile on the time-delay observed between source images~\citep{2007arXiv0706.3098G, 2013A&A...559A..37S, 2016JCAP...08..020B, 2016MNRAS.456..739X, 2021MNRAS.504.2224L}.

Data quality issues can bias the measurements. Single fiber spectroscopy introduces a number of biases, such as off-center fiber alignment, and suboptimal fiber exposure times which have effects on the pipeline determination of kinematic parameters, such as the galaxy velocity dispersion~\cite{2012ApJ...744...41B, 2012AJ....143...90S}. In contrast,~\cite{2018Sci...360.1342C} compared the dynamic mass measurement of spatially resolved stellar kinematics to the lens mass to find both agree well with General relativity. Modelling assumptions may also bias the measurements. In particular,~\cite{2018ApJ...855...64L} acknowledged that dynamic models that ignore the anisotropy of the system may introduce a bias. In addition, simple virial and Jeans derived dynamic mass estimators~\citep[such as][]{2009ApJ...704.1274W, 2010ApJ...710..886W, 2010MNRAS.406.1220W} typically assumed a virialized spherical galaxy with a velocity dispersion often assumed isothermal for massive galaxies and sampled from high SN spectra observed by a specific aperture. However, galaxies and available data rarely fit all of these qualifications. Often the aperture of the sample must be carefully accounted for to prevent or reduce biases~\citep{1980A&A....91..122M, 1981MNRAS.194..195B, 1983ApJ...266...58T}. One may also need to add a non-negligible surface term to the viral theorem~\citep{1986AJ.....92.1248T}. Previous attempts revealed there is often uncertainty in the measurement of the sample or selected galaxy isophote radius that must be carefully considered~\citep{2009ApJS..182..216K, 2010ApJS..191....1C, 2011A&A...533A.124G} to relate the stellar velocity dispersion to the gravitational potential. Newer estimators~\citep{2006MNRAS.366.1126C, 2013MNRAS.432.1709C} demonstrate reduced biases when isotropy is not assumed.

Alternatives to General Relativity, including Modified Newtonian Dynamics~\citep{1983ApJ...270..365M, 1983ApJ...270..371M, 1983ApJ...270..384M} and Modified Gravity theories~\citep{2006ApJ...636..721B,2019MNRAS.482.4514R} provide a possible resolution to the lens-dynamic mass discrepancy, since the weak-field regime of gravity is stronger than predicted by General relativity on scales larger than galaxies.  ~\cite{2017JCAP...08..036H} suggest that dark energy, which may add a concave lensing effect, may be a factor in the lens-dynamic mass discrepancy, in contrast to the hypothesis from~\cite{2011ICRC....5..223S}, that the cosmological constant has limited effects on lensing.

Lenses with sufficiently low redshifts of $z \lessapprox 0.1$ allow measurements of the stellar kinematics across the galaxy, which are ideal for testing possible causes of the lens-dynamic mass discrepancy for the following reasons:
\begin{itemize}
\item Spatially resolved kinematics can be used to reduce the uncertainties in the dynamic mass measurement. If the improved dynamic mass measurement resolves the lens-dynamic mass discrepancy, then low redshift lenses can be used to determine the bias in the dynamic mass modelling for galaxies at cosmological distances.
\item If improvements in dynamical modelling do not resolve the lens-dynamic mass discrepancy, then the discrepancy can be compared between low and high redshift lenses to search for possible explanations including differences in the internal galaxy structure or influences in the LOS density.
\item It is also possible that the lens-dynamic mass discrepancy may scale with the mass of the lens or environment. Any dependence of the discrepancy with lens mass may infer that the fitted lens model does not account for more mass than expected in the outer regions of the dark matter (DM) halo, which component would be measured in the lens mass along the LOS but less likely by stellar dynamics that measure the mass enclosed within stellar orbits.
\end{itemize}

Unfortunately, lenses are rare since the angular alignment of the source and lens must be of the order of an arcsecond relative to the observer. In addition, the angular position of low redshift source images likely resides within the foreground light of the lens. The best method to find samples of low redshift lenses is to spectroscopically detect emission-lines of star-forming sources within target spectra that the foreground has been modelled and subtracted, which method has yielded $\approx200$ intermediate redshift lenses from previous lens searches~\citep{2006ApJ...638..703B, 2011MNRAS.417.1601T, 2012ApJ...744...41B, 2015ApJ...803...71S, 2016ApJ...824...86S} within several sub-surveys of the Sloan Digital Sky Survey~\citep[SDSS;][]{2000AJ....120.1579Y}.

The Spectroscopic Identification of Lensing Objects~\citep[SILO;][]{2018MNRAS.477..195T, 2020arXiv200709006T} project is based upon the spectroscopic detection and selection methods of the Boss Emission-Line Lens Survey~\citep[BELLS;][]{2012ApJ...744...41B} to find on the order of $10^3$ lenses across a broadened range of redshifts, physical probe Einstein radii, and lens mass to statistically constrain the evolution in the galactic mass profile inferred by previous surveys. In this project, we found 1,551 lens candidates within the two million spectra of higher redshift galaxies contained within the completed Baryon Oscillation Spectroscopic Survey~\citep[BOSS;][]{2013AJ....145...10D} and its extended program~\citep[eBOSS;][]{Dawson_2016}, with intent to use lenses from this sample to statistically constrain extragalactic mass evolution. In addition, we initially scanned row-stacked-spectra (RSS)~\citep{2017AJ....154...86W}, which are individual fiber exposures from a fiber bundle pointed on each target, from the Mapping of Nearby Galaxies at Apache Point Observatory~\citep[MaNGA;][]{2015ApJ...798....7B} survey released in the fourteenth data release~\citep[DR14;][]{2017arXiv170709322A} of the fourth phase of the Sloan Digital Sky Survey~\citep[SDSS-IV;][]{2017AJ....154...28B}. The scan yielded 38 lens candidates~\citep{2018MNRAS.477..195T} from 2,812 low-redshift MaNGA galaxies, which warranted a scan of 10,000 galaxies within the completed MaNGA survey released in DR17~\citep{2022ApJS..259...35A} to use properties of the kinematic mapping and expected properties of the lens sample (see Section~\ref{data}) to resolve the cause of the lens-dynamic mass discrepancy and statistically constrain extragalactic mass evolution (see Section~\ref{intent}).

We scanned the co-added foreground subtracted RSS, which are stacked across exposures from the same fibre and dither position. The stacking enabled a sigma-clipping method to reject contamination from cosmic rays, satellites, or any other transient that typically affects a single exposure.

SILO also computes the spectroscopic redshift for each spectrum in MaNGA targets for use in modelling and subtraction of the foreground flux. These precise redshifts are being released in a spectroscopic redshift (Specz) value added catalogue (VAC) within DR17 (see Section~\ref{speczvac}). SILO also creates narrow-band images of the signal-to-noise (SN) of candidate background galaxies to enable a manual inspection of the spatial distribution of the signal to identify any lensed source features. The SN narrow-band images are extracted from spaxels we constructed from the co-added foreground subtracted RSS in order to mitigate signatures from transients, noise, and reduction issues. The narrow-band data is included in the lensing VAC released with DR17 (see Section~\ref{lensvac}).

This paper is organized as follows. The spectroscopic MaNGA data is described in Section~\ref{data}. How the SILO project plans to use this data is described in Section~\ref{intent}. The spectroscopic selection method is described in Section~\ref{ch4specsel}. Section~\ref{ch4results} describes the results. Sections~\ref{speczvac} and~\ref{lensvac} describe the Specz VAC and the lens VAC, respectively. Section~\ref{ch4conclusions} concludes with a summary and comments on the findings.

\section{Spectroscopic Data}\label{data}

The Mapping of Nearby Galaxies at Apache Point Observatory~\citep[MaNGA;][]{2015ApJ...798....7B} survey from the fourth phase of the Sloan Digital Sky Survey~\cite[SDSS-IV;][]{2017AJ....154...28B} has completed its observations of $\approx10,000$ low redshift galaxies. The completed observations of MaNGA galaxies included in the seventeenth data release of SDSS-IV \cite[DR17;][]{2022ApJS..259...35A} are based on a target selection function which yielded a volume-limited and fully representative sample of the local galaxy population around $z\approx0.03$, including an approximately flat distribution of targets across stellar masses greater than $10^8~M_{\odot}$, an approximately flat distribution of targets across absolute i-band magnitude, and no cuts on size, inclination, morphology, or environment \citep{2015ApJ...798....7B, 2017AJ....154...86W}. Thus MaNGA galaxies are ideal for finding a sample of low redshift lenses with various sizes, masses, morphologies, and environments across galaxy types larger than dwarfs. In addition, MaNGA uses an Integral-Field-Unit (IFU) to sample the spectra across the surface of the galaxy, which enables spatially resolved kinematic modelling of MaNGA targets~\citep{2015AJ....149...77D, 2016AJ....151....8Y, 2016AJ....152..197Y}. Each IFU is a bundle containing between 19 and 127 fibres, which the angular radius of each fibre is $1\arcsec$. The light collected by the 2.5 meter telescope at Apache Point Observatory~\citep{2006AJ....131.2332G} from each fibre is fed to the BOSS Spectrograph~\citep{2013AJ....146...32S}. The spectra obtained from each fibre and each 15 minute exposure are recorded in a row-stacked-spectra (RSS) format~\citep{2017AJ....154...86W}.

The MaNGA data reduction pipeline~\citep[DRP;][]{2016AJ....152...83L} also constructs a grid of $0.5\arcsec$ spaxels  in a 3-dimensional CUBE. Each spaxel is a stack of all RSS within $1.6\arcsec$, in which a flux-conserving variant of Shepard's method is used to apply an inverse-distance weight to the contribution from each RSS~\citep{10.1145/800186.810616, 2012A&A...538A...8S}.

\section{Aims and Objectives}\label{intent}

The discovery of the lens-dynamic mass discrepancy and the completion of the MaNGA survey warranted a scan of all MaNGA galaxies to obtain a sufficient sample of low-redshift lenses. Once obtained, this sample can be used to:
\begin{itemize}
\item Combine improved dynamic measurements to test the cause of the lens-dynamic mass discrepancy.
\item Probe the galactic mass profile around the baryon-dominated bulge of the lens since the strong lensing regime is projected to be approximately the physical size of the galactic bulge, which is small compared to the MaNGA target galaxy's angular size $\approx10\arcsec$~\citep{2018MNRAS.477..195T}, enabling evolutionary studies of the mass profile to be extended to lower physical radii.
\item Better constrain the initial mass function (IMF) using a sub-sample of lenses with small Einstein radii within the baryon-dominated galactic core, where dark matter does not significantly contribute to the gravitational mass, allowing lens models to constrain the IMF with reduced degeneracies~\citep{2013MNRAS.428.3183D, 2013MNRAS.434.1964S, 2015MNRAS.449.3441S, 2018MNRAS.478.1595C, 2020ARA&A..58..577S}.
\item Compare low-redshfit luminous-red-galaxy (LRG) lenses with intermediate redshift LRG lenses previously found within SDSS surveys to test the evolution in the galactic mass profile across a broadened range of redshifts.
\item Statistically infer the mass evolution in galaxy mass profiles at low redshifts if the lens sample size and mass range is sufficient.
\item Compare the 20 emission-line galaxy (ELG) lenses found by the Sloan WFC Edge-on Late-type Lens Survey~\citep[SWELLS;][]{2011MNRAS.417.1601T}, including potentially $\approx50$ ELG lenses (once confirmed) within SILO candidates found within the BOSS and eBOSS surveys, to a small potential sample of ELG lenses found within MaNGA, which boost in statistical power of the combined lens sample and improved dynamic information of low-redshift lenses can help break the bulge, disk, and halo mass components statistically inferred at intermediate redshifts~\citep{2011MNRAS.417.1601T}.
\end{itemize}

\section{Spectroscopic Selection}\label{ch4specsel}

The spectroscopic selection method applied to find lenses within the completed MaNGA survey of DR17 is based upon the methods in~\cite{2012ApJ...744...41B, 2018MNRAS.477..195T, 2020arXiv200709006T}. The first stage in this spectroscopic selection process is to apply the BELLS spectroscopic detection method~\citep{2012ApJ...744...41B} to isolate high SN candidate background emission-lines embedded within foreground galaxy spectra obtained from single-fiber observations, which is summarized in two main steps:
\begin{enumerate}
    \item{\bf Foreground galaxy subtraction.} Model and remove the foreground galaxy spectra to isolate background emission-lines within the residuals (see Section~\ref{ch4foreground}). The fitting process requires a precise redshift to properly scale the wavelength of the galaxy spectral templates to be fitted to the flux. Since MaNGA data only contains an overall galaxy redshift, the SILO software also computes the spectroscopic redshift for each fiber (see Section~\ref{speczcomp}) for use in foreground modeling and subtraction.
    \item {\bf Search for remaining background signals.} Scan for sets of high SN signals within co-added residuals (see Section~\ref{coaddition}) with wavelength separations that match typically observed sets of background emission-lines. An automated BELLS pre-inspection cut next rejects detections more likely explained as contamination prior to the manual inspection process. Inspectors then manually identify which detections demonstrate flux patterns of expected background emissions (see Section~\ref{ch4detect}).
\end{enumerate}

MaNGA targets are observed through fiber bundles with angular coverage of $\approx10\arcsec$, enabling examination of the spatial location of the candidate background signal to identify strong lensing features, background galaxies too far away to be strongly lensed, or if the signal is a form of foreground or intervening contamination. Thus extra steps were added~\citep{2018MNRAS.477..195T} to the spectroscopic selection process to utilize the spatial information of MaNGA targets:
\begin{itemize}
    \item {\bf Compute the probable strong lensing regime.} An upper limit projection of the Einstein radius is computed to determine if the spatial locations of the detected signal reside within the probable strong lensing regime (see Section~\ref{ch4er}).
    \item {\bf Construct a narrowband image of detected emissions.} Narrow-band spaxel images of the candidate background emission-lines are constructed to reveal the detected signal demonstrated lensing features, a background galaxy, or a form of contamination (see Section~\ref{ch4narrowband}).
    \item {\bf Identify potential lensing features.} Candidate source-planes that contain assuring detection(s) near the same redshift are inspected to identify whether the spatial distribution of the signal across the narrow-band image and the detections contain identifiable lensing features, including a precise spectroscopic redshift for a candidate source within the probable strong lensing regime, or indications of contamination from either the foreground galaxy or any intervening objects (see Section~\ref{ch4sourceplane}).
\end{itemize}

However, instead of scanning the foreground subtracted RSS (i.e., residuals) as in~\cite{2018MNRAS.477..195T}, SILO now scans the co-added residuals stacked across exposures from the same fibre at the same dither position so a sigma-clipping method can reject transient signals. The co-addition of the residuals reduces the number of false-positive detections.

The narrow-band images are also now constructed from the co-added SN, which provides several advantages over conventional narrow-band image construction from foreground subtracted MaNGA spaxels. The first advantage is SN narrow-band images suppress noisy regions relative to narrow-band images of residuals. Spaxels constructed from co-added residuals are also free of false-positive signals caused by RSS masking, which issue is caused by constructing the affected spaxel from a set of RSS with a stacked flux-density that is greater than one of the RSS in the stack that is masked at the wavelength region of the constructed narrow-band. This section also describes the improvements and the inclusion of new tools from~\cite{2020arXiv200709006T} to the spectroscopic detection, spectroscopic selection, and source-plane inspection methods. We describe the spectroscopic selection process in the following subsections, including the method used to determine the spectroscopic redshift for each spectrum (see Section~\ref{speczcomp}), the use of the redshift to subtract the foreground flux from each RSS (see Section~\ref{ch4foreground}), the removal of transient signals using co-added residuals (see Section~\ref{coaddition}), the search for background emission-lines (see Section~\ref{ch4detect}), setting bounds for the probable strong lensing regime (see Section~\ref{ch4er}), generating narrow-band images of each background detection across the geometry of the fiber bundles (see Section~\ref{ch4narrowband}), and the manual inspection of the narrow-band images and proximity of the detection to the probable strong lensing regime in order to identify promising lens candidates (see Section~\ref{ch4sourceplane}).

\subsection{Computation of the Spectroscopic Redshifts}\label{speczcomp}

As in~\cite{2018MNRAS.477..195T}:
\begin{itemize}
    \item The {\sc spec1d -- zfind} code from the publicly available BOSS pipeline~\citep[][]{2012AJ....144..144B} is used to perform a principal component analysis (PCA) fit of the spectra to determine the spectroscopic redshift of the fit, in which the redshift from the NASA Sloan Atlas~\cite[NSA;][]{2017ApJS..233...25A} catalogue is used as the initial guess in the first pass.
    \item SILO next measures the mean of the computed spectroscopic redshifts located within the high SN inner region of the galaxy.
    \item SILO then uses the {\sc spec1d -- zfind} software in a second pass of the computation of the spectroscopic redshifts, in which the mean redshift is used as the initial guess to improve the evaluation of spectra at low SN. The MaNGA RSS and MaNGA spaxel spectra are evaluated separately.
\end{itemize}

The method to select the high SN inner region of galaxies is replaced with a new method based on identifying the radius where all enclosed redshifts are measured within the LOS velocity dispersion of the galaxy. The steps to compute the high SN inner region of the galaxy are described in the following:
\begin{itemize}
\item SILO rejects unphysical redshifts with a relativistic Doppler shift $\pm 1,000\ \mathrm{km\,s}^{-1}$ from the innermost redshifts to the galaxy centre.
\item SILO next uses a maximum-likelihood-estimation (MLE) to fit a Gaussian with a uniform background to the sample of remaining redshifts to compute the LOS dispersion of the redshifts ($\sigma_{MLE}$). The dispersion is caused by the relativistic Doppler shifts induced by the rotational velocity field of the galaxy.
\item The inner high SN radius ($R_{inner}$) is set as the maximum radius no spectroscopic redshift diverges more than three times the dispersion of the redshifts from the median of the enclosed sample.
\item SILO then computes the mean ($\mu_{inner}$) and the standard deviation ($\sigma_{inner}$) of the redshifts within $R_{inner}$.
\item SILO also computes the mean galaxy redshift from all redshifts within three $\sigma_{MLE}$ of $\mu_{inner}$.
\end{itemize}

\begin{figure*}
\begin{center}
\subfloat[Row Stacked Spectra]{\includegraphics[height=7.2cm]{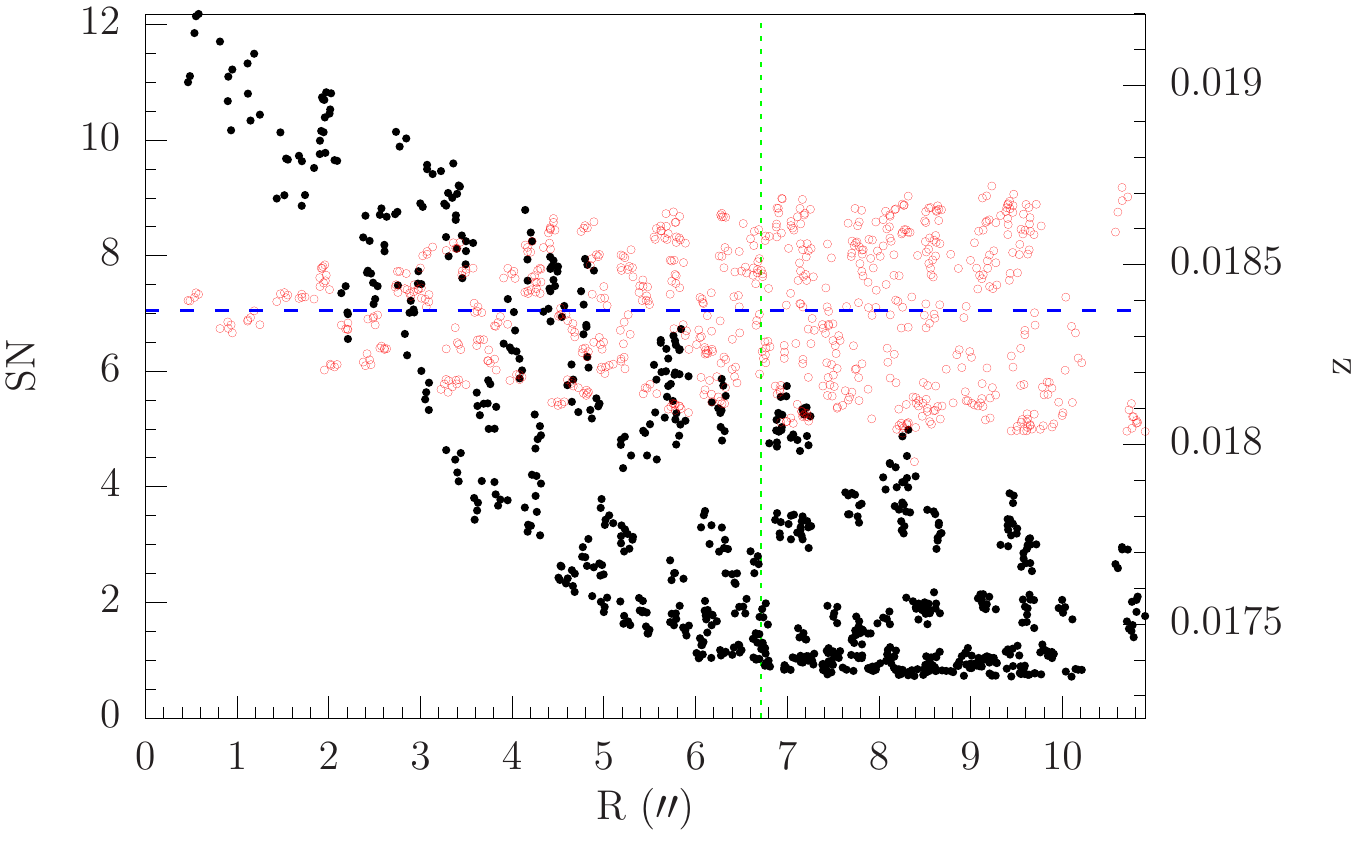}
\includegraphics[height=7.2cm]{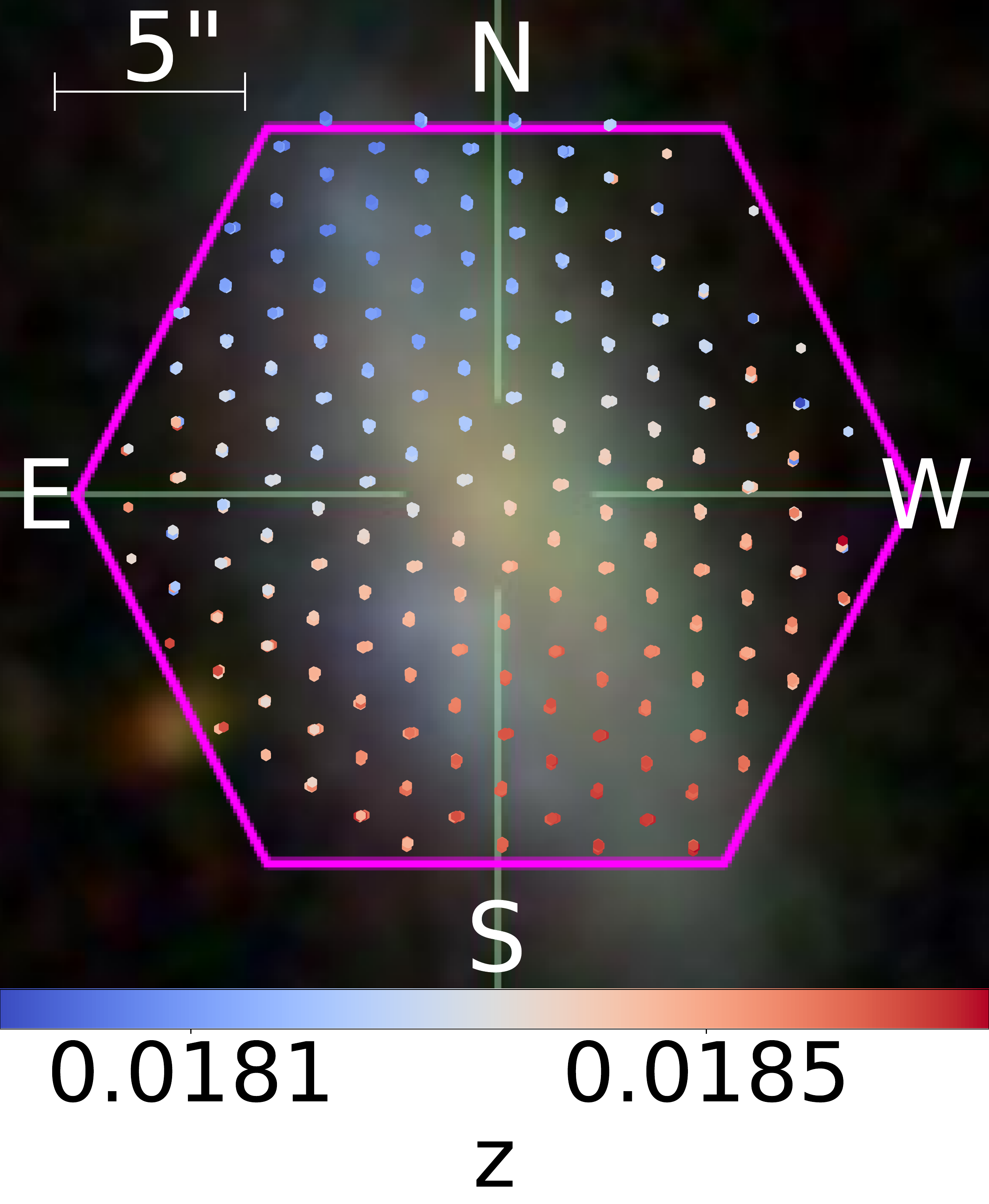}}
\end{center}
\begin{center}
\subfloat[CUBE Spaxels]{\includegraphics[height=7.2cm]{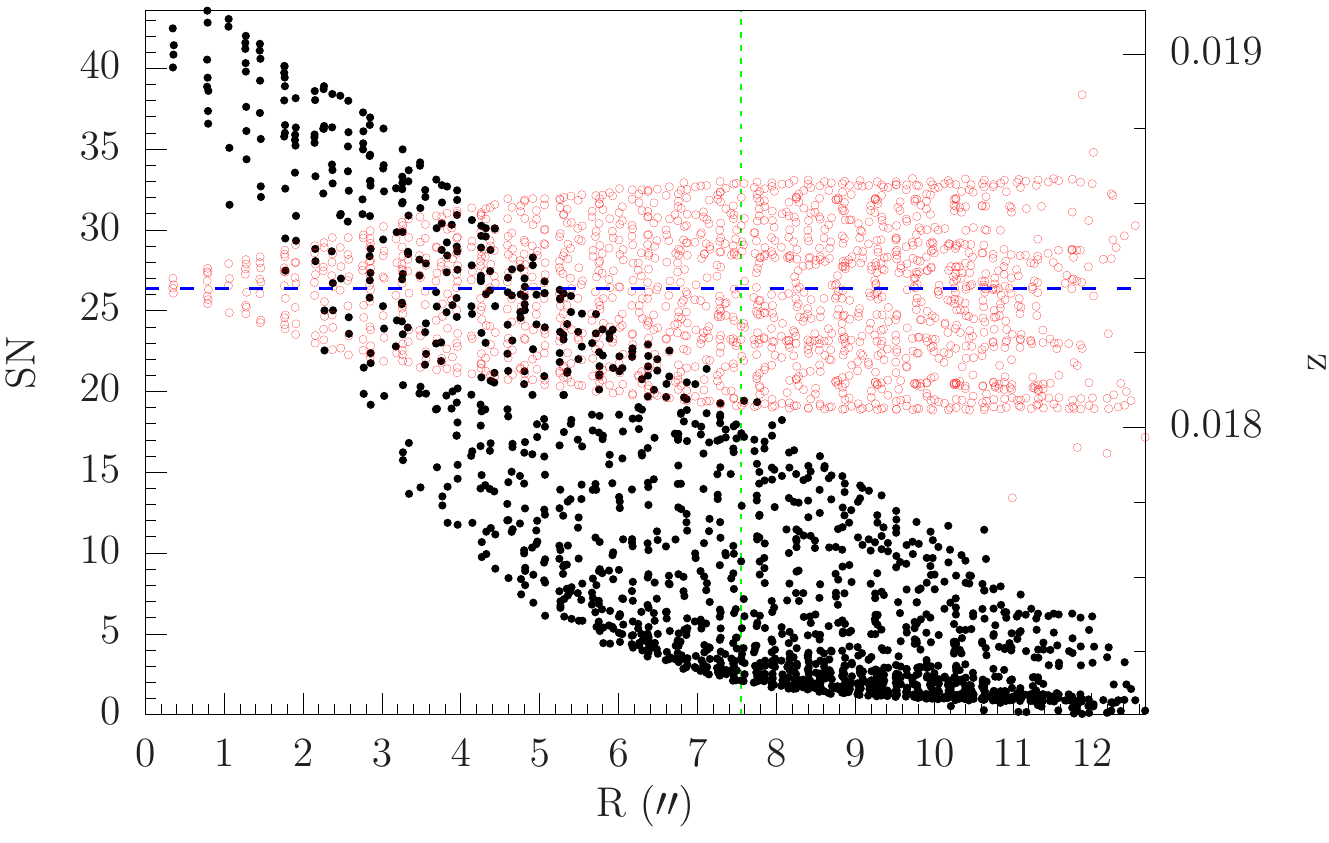}
\hspace{4px}
\includegraphics[height=7.2cm]{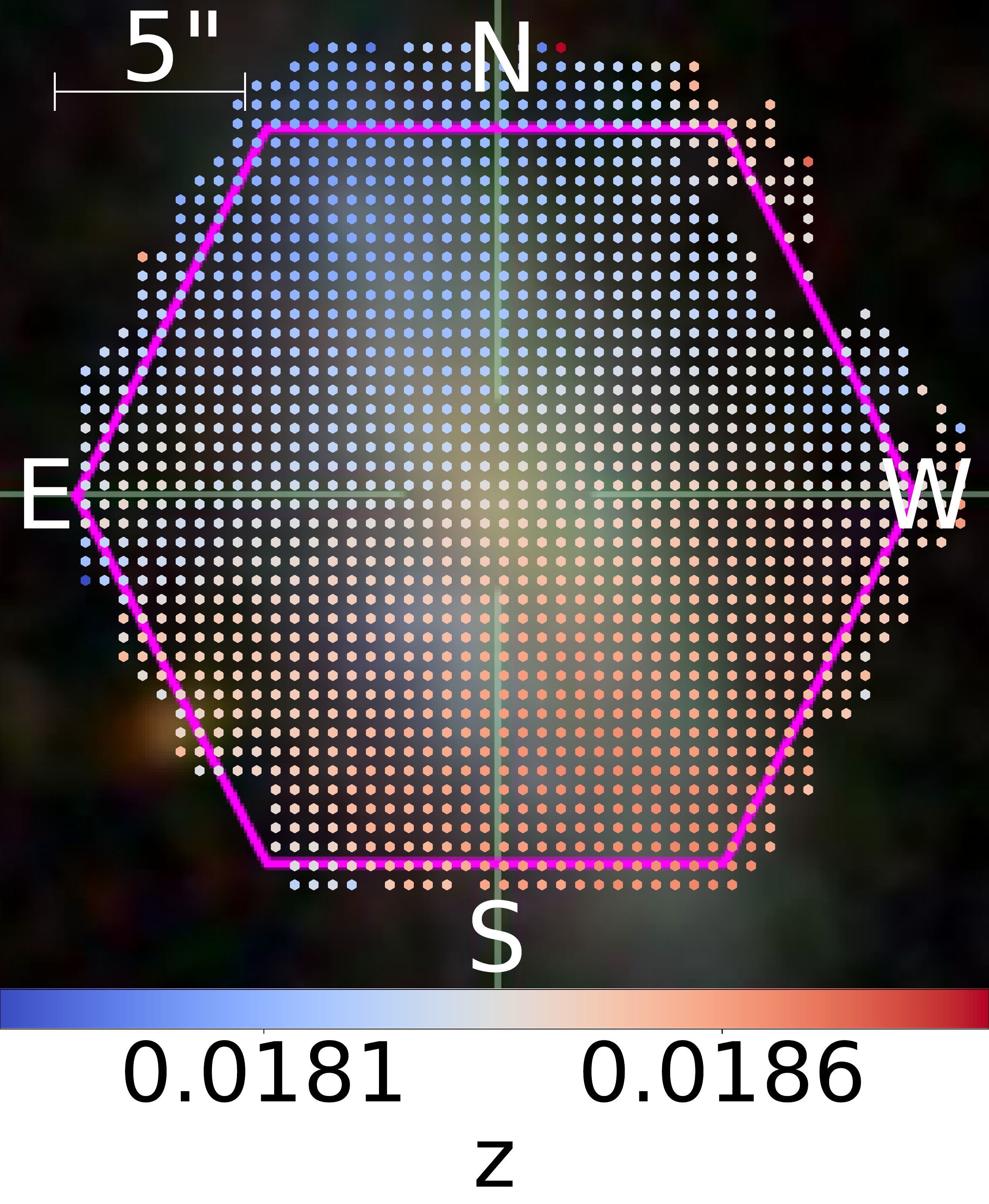}}
\end{center}
\caption[Examples of spectroscopic redshifts contained within the Specz VAC]{RSS (top) and CUBE spaxel (bottom) examples of spectroscopic redshifts contained within the Specz VAC. Left: Each plot displays the radial distribution of spectroscopic redshifts (red) computed for the spectra from galaxy SDSS J1525+4310 with sufficient SN (black) to model. The dotted green vertical line represents the high SN inner radius evaluated for each galaxy and defined in Section~\ref{speczcomp}. The dashed blue horizontal line represents the mean redshift evaluated from the spectroscopic redshifts enclosed within the high SN inner radius. Both RSS and CUBE spaxel images are contained in the Specz VAC, as described in Section~\ref{speczvac}. Right: Spatial locations of the spectroscopic redshifts are marked with circles and overlayed on the SDSS target image. The color represents the redshift specified in the colorbar.}
    \label{fig:speczexample}
\end{figure*}

The spectroscopic redshifts that are within three $\sigma_{MLE}$ of $\mu_{inner}$ from the second pass are presented in the Specz VAC for a selection of targets as described in Section~\ref{speczvac}. Figure~\ref{fig:speczexample} demonstrates the radial distributions of the spectroscopic redshifts for both RSS and spaxel reductions for a representative galaxy, SDSS J1525+4310, and the specz values are increasingly correlated with the NSA (reference) redshift toward galactic center. The relativistic doppler shift in each spectroscopic redshift is apparent in Figure~\ref{fig:speczexample} and is caused by the LOS velocity profile of the galaxy.

\subsection{Foreground Galaxy Subtraction}\label{ch4foreground}

As in~\cite{2018MNRAS.477..195T}, the computed spectroscopic redshifts are used as input in a seven-component PCA fit of galaxy eigenspectra to the foreground flux.  The fit uses a more complete basis of seven-component eigenspectra created from a PCA decomposition of a sample of hundreds of galaxy spectra within the SDSS survey~\citep{2011ApJS..193...29A}, rather than the four PCA eigenspectra used in the pipeline's redshift analysis.  This provides a better fitting model, mathematically possible because the redshift is utilized as a prior.   The best-fit is then used to subtract the foreground flux. The foreground subtraction leaves a noise-dominated residual spectrum that can be searched for background emission-lines.

\subsection{Co-Addition of Residual Spectra}\label{coaddition}

Stacking the foreground-subtracted RSS at the same position on the sky allows cosmic rays and other transient contaminants to be identified and removed since these signals are not present in all stacked spectra. The SILO software uses the \texttt{combine1fiber} function from the \texttt{spec2d} package within \texttt{PYDL}\footnote{PYDL: \url{https://pypi.org/project/pydl}}~\citep{zenodo.1095151} to co-add foreground-subtracted residuals across exposures from the same fiber taken at the same dither position so \texttt{combine1fiber} can use a sigma-clipping method to reject transients.

\subsection{Background Emission-Line Detection}\label{ch4detect}

The spectroscopic detection method is designed to discover emission-lines from background galaxies embedded in the foreground galaxy spectra, provided the redshifts of the background emission-lines are within the observed wavelength range of the spectrograph. Since the wavelength window of the BOSS spectrograph is 3,600\AA -- 10,400\AA~\citep{2013AJ....146...32S}, the detection limit of each emission-line has a maximum redshift, $z_{max}$, as listed in Table~\ref{table:emlines}. These background emission-lines are not modelled in the SDSS pipeline fit of the foreground flux and thus remain in the residuals. The SILO spectroscopic detection method described in~\cite{2018MNRAS.477..195T} is applied to each co-added foreground-subtracted residuals to detect candidate $[O\,\textsc{ii}]$(b, a) with a $SN\ge6$ (single-line) or multiple emission-lines from Table~\ref{table:emlines} with a $SN\ge4$ (multi-line), in which detected $[O\,\textsc{ii}]$(b, a) doublets are counted as a single emission-line during the multi-line scan. These scans yielded 80,954 multi-line and 33,738 single-line detections from 1,247,568 co-added residuals stacked across 5,402,098 foreground-subtracted RSS, which RSS is obtained from 11,980 MaNGA DRP reductions of the $\approx10,000$ MaNGA target galaxies. The candidate emission-lines are then fitted to Gaussians with the same method as previously applied in~\cite{2018MNRAS.477..195T} and~\cite{2020arXiv200709006T}.

\begin{table}
\caption{Emission-lines searched by SILO. The listed emission-lines were used to scan for background galaxy candidates. Column one lists the name of each emission-line. Column two lists the wavelength in a vacuum of a restframe. Column three lists the maximum redshift each emission-line can be detected by the BOSS spectrograph.}
\centering
\label{table:emlines}
\setlength\tabcolsep{4pt}
\begin{tabular}{lcc}
 \hline\hline
Emission & Restframe & {\(z_{\mathrm{max}}\)} \\
Line & Wavelength [\AA] & \\
(1) & (2) & (3)\\
\hline
$[O\,\textsc{ii}]$b & 3727.09 & 1.78 \\
$[O\,\textsc{ii}]$a & 3729.88 & 1.78 \\
H${\delta}$ & 4102.89 & 1.52 \\
H${\gamma}$ & 4341.68 & 1.38 \\
H${\beta}$ & 4862.68 & 1.13 \\
$[O\,\textsc{iii}]$b & 4960.30 & 1.09 \\
$[O\,\textsc{iii}]$a & 5008.24 & 1.07 \\
$[N\,\textsc{ii}]$b & 6549.86 & 0.58 \\
H${\alpha}$ & 6564.61 & 0.58 \\
$[N\,\textsc{ii}]$a & 6585.27 & 0.57 \\
$[S\,\textsc{ii}]$b & 6718.29 & 0.54 \\
$[S\,\textsc{ii}]$a & 6732.68 & 0.54  \\
\hline
\end{tabular}
\end{table}

The SILO software then applies pre-inspection cuts that are similar to those used in ~\cite{2012ApJ...744...41B, 2020arXiv200709006T} to reject signals that are more likely explained as sky emissions, foreground emissions, or misidentified emissions. We briefly describe each step of the pre-inspection cut, followed by any differences in the method between the single-fibre detection methods applied in~\cite{2020arXiv200709006T} and the BELLS survey and our scan of MaNGA IFUs.

The first step of the pre-inspection cut is to use histograms of the detections across the observed and restframe wavelengths to identify where binned counts are greater than the sliding median by a threshold. The threshold is adjusted by the inspector until only the observed regions of contamination are masked. In~\cite{2020arXiv200709006T}, all single-line detections whose candidate $[O\,\textsc{ii}]$(b, a) doublet is near in wavelength to an unusually high occurrence of detections in either the rest-frame or observed-frame are rejected as target or sky emission-lines, respectively. However, many $[O\,\textsc{ii}]$(b, a) doublets detected within an IFU spectra from the same background galaxy may be misidentified as sky or target emissions within the histogram method. Thus only one detection at a specific redshift and for a specific target is used in the histogram method.

As in~\cite{2020arXiv200709006T}, single-line detections are next rejected in the pre-inspection cut if a multi-line detection also exists at the same redshift within the co-added spectra. Single-line detections are also rejected if a candidate $[O\,\textsc{ii}]$(b, a), H${\delta}$, $[O\,\textsc{iii}]$a, or H${\alpha}$ emission-line with SN $\geq 3$ is present when treating the single-line detection as a misidentified H${\delta}$, $[O\,\textsc{iii}]$a, or H${\alpha}$ emission-line. Only 7,158 multi-line and 1,124 single line detections passed the pre-inspection cuts.

In the pre-inspection cuts applied by~\cite{2012ApJ...744...41B} and the BELLS survey, detections whose lens-source redshifts produce a test Einstein radius less than 4.5 kiloparsecs were rejected under the suspicion that strong lensing may not exist for the system. This is because any applied Singular Isothermal Ellipsoid~\citep[SIE;][]{1994A&A...284..285K} with a test mass that would generate a fiducial 250 km\,s$^{-1}$ velocity dispersion within an SDSS fibre will likely over project the mass enclosed at smaller radii. This pre-inspection cut is not applied to MaNGA lens candidates since spatially sampled stellar and dynamic information is available to apply a more detailed analysis of the mass enclosed at smaller radii.

The manual inspection process used to grade candidate background emission-lines is necessary in order to label the visible features including a subjective quality assessment.  Our method is nearly identical to the steps described in Section 3.5 of \cite{2020arXiv200709006T}, which is mostly unchanged but with some additional refinements, from the manual inspection process used by \citet{2012ApJ...744...41B}. In summary, detections are manually inspected in an initial pass by a single individual, labeling visible emission-line patterns, which work to increase the confidence in the grade, and labelling any reduction artifacts, sky lines, foreground emission-lines, or nearby intervening objects, which work to decrease the confidence in the grade.  A second pass is then made including a second individual in order to ensure the grading is robust by ensuring a more uniform analysis.  The results of the manual inspection are finally converted into a simple letter-grade, in which candidates with a Grade A represent likely, Grade B represent probable, Grade C represent possible, and Grade X represent doubtful graviational lens candidates.  The advantage of providing the final coursely grained letter-grade is to minimize the subjective nature of the manual inspection process.  However, we are currently testing machine learning methodologies to either supplement or replace the manual inspection process.

\begin{figure*}
    \begin{center}
    \includegraphics[scale=.78]{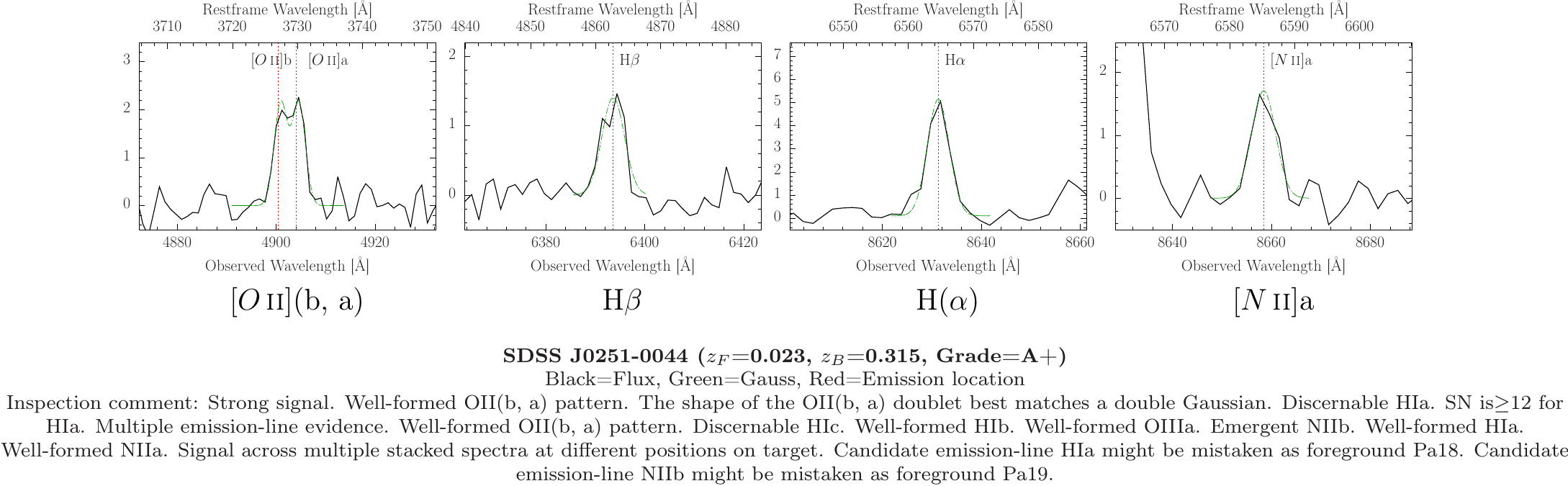}
    \caption[Examples of candidate emission lines]{Examples of candidate emission lines. This plot displays the $SN\ge4$ candidate background emission lines for a spectroscopic detection within the co-added residuals (black line). The Gaussian fitted to each emission line (green dash) is overplotted. Red dotted lines represent the locations of each emission line. The target name, lens redshift, source redshift, inspection grade, VAC labels, and inspection comments are listed below the plot to preserve the format as seen in the lens VAC (see Section~\ref{lensvac}). The inspection comments may include descriptions of lower-SN candidate emission lines observed during the spectra inspection of the detection. The same descriptive sentence of the signal can appear twice in the inspection comments for each VAC plot, in which the first describes the quality of an emission-line pattern while the second is part of a list that supports the presence of multiple emission-lines.}
    \label{fig:emlines}
    \end{center}
\end{figure*}

In~\cite{2020arXiv200709006T}, the same detection observed across neighbouring fibres is flagged as a false positive since it is more likely that the signal occurs from sky contamination than from a detection at the same redshift within a different target galaxy. However, the MaNGA IFU spectra are stacked together, and thus neighbouring fibres do not enable comparison across different galaxies. Instead, a background galaxy can be assured with confidence by observing the same signal across a set of fibres spatially localized near the detection. The sky contamination can typically be identified by observing if the spatial distribution of the detected signal is not localized to a position on sky. The high SN emission-lines of one highly assured detection are demonstrated in Figure~\ref{fig:emlines}, which represents the typical quality of a Grade A+ (i.e. highly likely) detection.

\subsection{Einstein Radius Estimation}\label{ch4er}

Previous searches for galaxy-galaxy graviational lenses within SDSS, BOSS and eBOSS surveys used single-fiber spectra, and assumed that the observed background emission-lines were likely in the strong lensing regime because the Einstein radius was typically similar to the angular size of the fibre. However, MaNGA targets are nearby low-redshift galaxies, and therefore would typically have Einstein radii much larger than the single-fiber radius, possibly as high as $10\arcsec$ compared to the $1\arcsec$ Einstein radius of SLACS and BELLS lenses.  Fortunately, the MaNGA IFU provides many spectra across the 2 dimensional plane of the galaxy, from which spatially sampled stellar mass and stellar kinematic maps have been computed, and included in the data release.  We use these maps to compute a projection of the density profile of each MaNGA target so the probable strong lensing regime can be approximated as an upper limit to the Einstein radius (UER).

Uncertainties in the density profile introduce relative uncertainties in the projected Einstein radii, and thus this radius cannot be directly used as a rejection threshold without the risk of rejecting real lenses with underestimated projected Einstein radii.  As in~\cite{2018MNRAS.477..195T}, the UER is defined as the radius where the upper limit of the integrated mass is the same as the mass enclosed within the test Einstein radius, and is used as a rejection threshold minimizing the possibility of excluding real lenses.

Although the UER is useful to judge the likelihood that the candidate background galaxy may be within the strong lensing regime, the computation can be sensitive to uncertainties in stellar mass maps, especially if the average density within the central region of the galaxy is near the critical density for lensing. Thus the source-plane inspection (see Section~\ref{ch4sourceplane}) now includes a test of the robustness of the probable strong lensing regime based on the UER derived from different mass maps.

The combined effects of the seeing and reduction methods used to create a mass map can also smooth the central density profile, and thus the enclosed mass at small radii can be underestimated. To mitigate the effects of smoothing, SILO now fits a mass density model to the mass data, which method uses a point-spread function (PSF) to de-convolve the density model. This section outlines two different computations of the density models based upon either fitting the MaNGA kinematics or fitting stellar mass maps multiplied by a DM fraction (DMF). The former computation of the UER directly probes the galaxy's total mass profile and thus can be more accurate when quality issues are not present in the MaNGA kinematics. We assume a flat $\Lambda$CDM cosmology. Unless otherwise specified, the SILO code defaults to cosmological parameters specified by the nine-year Wilkinson Microwave Anisotropy Probe~\cite[WMAP;][]{2013ApJS..208...19H} with $H_0 = 69.3\ \mathrm{km\,s}^{-1}\mathrm{Mpc}^{-1}$ and $\Omega_m = 0.287$.

\subsubsection{Computation of the Total Density Map using Stellar Dynamics}\label{cjam}

We compute the total density of the galaxy by performing a best-fit of a density model to the stellar velocity maps of each MaNGA target
Though the individual orbits of stars are not resolved at cosmological distances, the spatially resolved features in the velocity maps can be used to constrain the orbital parameters so the magnitude of the stellar velocities can be better evaluated. For example, the LOS stellar velocity $v_{LOS}$ is the observed LOS component of the rotational velocity $V_{rot}$, and thus $v_{LOS}$ depends on the inclination and the position-dependent direction of rotation. The LOS stellar velocity dispersion $\sigma_{LOS}$ can also be used to infer the magnitude of a field of randomly directed stellar orbits ($\sigma$) if the anisotropy and its angle relative to the observer is known. Due to the variation in the magnitude of the velocities with the distance from the galaxy centre, both $V_{rot}$ and $\sigma$ imprint spatially dependent patterns onto the LOS velocity maps that can be modelled to reduce uncertainties in the magnitude of the velocity field.

The MaNGA Data Analysis Pipeline~\citep[DAP;][]{2019AJ....158..160B, 2019AJ....158..231W} used Shepard's method to construct $v_{LOS}$ and $\sigma_{LOS}$ stellar velocity maps. However, the DR14 stellar velocity maps contained global data quality issues and thus are not used in the first search for lenses within MaNGA targets of DR14~\citep[see][]{2018MNRAS.477..195T}. The stellar velocity dispersion measurements are biased high from systematics, and where it became difficult to measure the spectra broadening in low SN spectra~\citep{2019AJ....158..231W}. The DAP has mitigated both issues for MaNGA reductions being released in DR17. In particular, the DR17 reductions of MaNGA targets include a stellar velocity dispersion map evaluated from Voronoi bins~\citep{2003MNRAS.342..345C}, in which MaNGA stacked each spatial sector until an of $SN\ge10$ is achieved~\citep{2019AJ....158..231W}. The stellar velocity dispersions computed for sectors of the galaxy are presented in the \texttt{VOR10-MILESHC-MASTARSSP} files scheduled for release in DR17, by which \texttt{MASTARSSP} indicates the MaStar-based integrated spectra of simple stellar population (SSP) models~\citep{2020MNRAS.496.2962M}. The DAP also produced a systematics correction map~\citep{2019AJ....158..231W} to the stellar velocity dispersion map.

SILO uses a Jeans Anisotropic Modeling~\citep[JAM;][]{2008MNRAS.390...71C, 2020MNRAS.494.4819C} of the kinematics to fit a total density model. The construction and fit of the density model is described in Appendix~\ref{jamdynamics}. The upper limit of the total density model is used in a computation of the UER.

\subsubsection{Computation of Total Density Maps Using Stellar Mass Maps}\label{fireflyer}

Due to the quality issues within the MaNGA kinematic maps, it was decided in~\cite{2018MNRAS.477..195T} that the most accurate method to compute an upper limit of the density map was to multiply the upper mass limit of stellar-mass maps with a Navarro-Frenk-White~\citep[NFW;][]{1996ApJ...462..563N} DMF. To mitigate the UER from being inflated by stellar-mass uncertainties, the stellar-mass maps obtained from the MaNGA FIREFLY VAC~\citep{2017MNRAS.466.4731G, neumann} are used in~\cite{2018MNRAS.477..195T}, in which the relative errors of the mass maps are fractional to the stellar mass estimates.  FIREFLY is a chi-squared minimization code that computes best-fit combinations of the \texttt{M11-MILES} single-burst stellar population models~\citep[][]{2011MNRAS.418.2785M} to the spectral energy distribution for each MaNGA IFU spectrum. Thus uncertainties in FIREFLY mass measurements are significantly less than photometric measurements~\citep{2015MNRAS.452.3209R} that multiply the light of the galaxy with a stellar population assessment of the scale and gradient of the stellar mass to the stellar light ratio ($M_*/L_*$). \cite{2017MNRAS.466.4731G} assumes Planck cosmological constraints~\citep{2016A&A...594A..13P} and a \citet{2001MNRAS.322..231K} IMF. The difference between WMAP and Plank parameters introduces a bias that is insignificant to the overall stellar mass and NFW DMF uncertainties used to determine the upper limit of the strong lensing regime.

The galaxy light is used to fit the gradient in the stellar density model since the shape of the galaxy profile is better fitted by SDSS photometry than FIREFLY stellar mass maps due to the improved spatial resolution of the image that is not affected by MaNGA reconstruction processes. SILO constructs a stellar density model by fitting a model of the seeing convolved light to the target's photometry and then re-scale the light model to approximate the stellar surface density observed in FIREFLY maps. In particular, SILO fits the photometry from the SDSS-I/II Legacy survey~\citep{2001ASPC..238..269L}, which images are obtained from the SDSS camera~\citep{1998AJ....116.3040G}, with a Multi-Gaussian-Expansion (MGE) that is convolved by the image PSF (see Appendix~\ref{apdx_stellar_mass_profile}). Masked regions in photometry are ignored during the MGE fit, which masking is described in Appendix~\ref{apdx_stellar_mass_profile}. The radially-independent $M_*/L_*$ is next computed by the ratio of the modelled light to stellar mass enclosed within four times the reconstructed MaNGA PSF ($PSF_M$), which radius is chosen based on how close the modelled light can be scaled to the inner stellar mass of the galaxy without introducing a significant smoothing bias from MaNGA reconstruction processes. The position-dependent total stellar mass per voronoi bin are obtained from the \texttt{STELLAR\_MASS\_VORONOI} extension within the MaNGA FIREFLY VAC file. The light profile is then multiplied $M_*/L_*$ to approximate the stellar mass profile.

The radially-dependent NFW DMF is obtained from \cite{2015ApJ...799..149J}, with Bayesian measurements derived from a joint likelihood fit of the stellar to dark matter fraction and size of the emission-line region within the quasar to microlensing maps created for 27 quasar image pairs across 19 lenses. The fit is reasonable since microlensing is sensitive to low stellar mass fractions~\citep{international2004dark}. These uncertainties are significantly less than comparing stellar-mass approximations~\citep{2015MNRAS.452.3209R} to the lens mass. The upper limit of the total mass map is determined from the upper limit of the stellar mass maps multiplied by the radially dependent upper limit of the NFW DMF.

\subsection{Narrow-band SN Image Construction}\label{ch4narrowband}

In~\cite{2018MNRAS.477..195T}, narrow-band images of the residual flux were created as a tool to identify if a candidate background galaxy demonstrated lensed features. In particular, SILO integrated the foreground-subtracted spectra near the wavelength(s) of the background emission-line(s) within the MaNGA spaxels. SILO has since been upgraded to create narrow-band images from the SN of the background emission-line(s) observed within spaxels constructed from the SN of co-added residuals, which new method suppresses the following types of possible contamination (see Figure~\ref{fig:narrowbandcompare}):
\begin{itemize}
\item Spaxels inherit contamination from incompletely masked transients within the RSS. SILO spaxel construction from co-added residuals bypasses this issue since sigma-clipping rejects transients when the residuals are co-added across exposures (see Section~\ref{coaddition}).
\item Transient masking can induce a flux bias in spaxels constructed from neighboring RSS if the masked region in a neighboring RSS contains a mean flux that is relatively less than other neighboring RSS since Shepard's method will only measure the remaining un-masked and relatively higher fluxes at the affected wavelengths. The flux bias often appears near the galaxy's center since the relative flux between neighboring RSS is inflated due to the steep radial gradient in the galaxy light profile. The shape of the flux bias varies but can even resemble a source image in the shape of an arc. SILO spaxels constructed from co-added residuals do not inherit this flux bias since the residuals have a mean flux of zero.
\item Improperly subtracted sky, non-perfect subtraction of brighter regions of galaxies, or other forms of non-transient contamination within the RSS can induce a bias in spaxels. These affected regions are often related to fitting spectra with quality issues and higher uncertainty measurements. Thus remaining forms of contamination are often suppressed in SILO SN narrow-band images.
\end{itemize}
In particular, SILO applies the method used to construct MaNGA spaxels to create SILO spaxels of the SN across the full wavelength range of the co-added residuals, which the single-line and $[O\,\textsc{ii}](b, a)$ doublet SN vectors are computed during the spectroscopic scan for background emission-lines. The narrow-band images are finally created by plotting the SN for each spaxel, which SN is added in quadrature for all detected emission-lines. However, it is important to note that the SN suppresses background signals with a flux less than the noise, depending on the radius from the target centre.

\subsection{Source-Plane Inspection}\label{ch4sourceplane}

The objective of the source-plane inspection is the same as in~\cite{2018MNRAS.477..195T}, which is to determine if the quality, spatial location of the detection(s), and the features of the signal observed in narrow-band images assure a background galaxy is real and likely lensed. However, the source-plane inspection now includes a grading scheme customized to inspect detections from co-added residuals instead of individual residuals. The grading scheme to gauge the assurance of the candidate source is described in Table~\ref{table:source_grading}. Rules developed in~\cite{2018MNRAS.477..195T} to reject background galaxy candidates that contained only a single-line detection do not apply to the detections within the co-added residuals since the signal observed across exposures is not likely caused by transient contamination.

As in~\cite{2018MNRAS.477..195T}, the manually inspected grade of a candidate background galaxy can be upgraded if the signal resides within twice the UER to reflect the increased confidence of the detection within the strong-lensing regime. However, the improvements in the DAP stellar velocity maps enable a second computation of the UER to test the robustness of the candidate strong lensing regime across several computation methods. The grading scheme described in Table~\ref{table:source_grading} is designed to test the likelihood that the candidate background galaxy is lensed based upon the strength and robustness of the candidate strong lensing regime, the proximity of the candidate background galaxy to the candidate strong lensing regime, the quality of the SDSS photometry and velocity maps, and any lensed source features within the narrow-band images. In the case that JAM-based UER computations are significantly larger than typical Einstein radii projections, which is caused when JAM fitting uncertainties are inflated by the resolution of two or more local $\chi^2$ minima regions, tThe inspector will instead compare the proximity of detections tohe the next best upper measurement of the Einstein radii projection.

\section{Results}\label{ch4results}

This section summarizes the results in the same order as the spectroscopic selection process. In particular, results from the redshift computations, the spectroscopic scan, inspection of candidate emission signals, computations of the ER, comparison of SN to flux narrowband images, and inspection of candidate sources are divided into the following sub-sections. All corner plots represent distributions of fitted model parameters across samples specified in the following sections or figure captions.

\subsection{Spectroscopic Redshifts Computations}

\begin{figure*}
\begin{center}
\includegraphics[width=\textwidth]{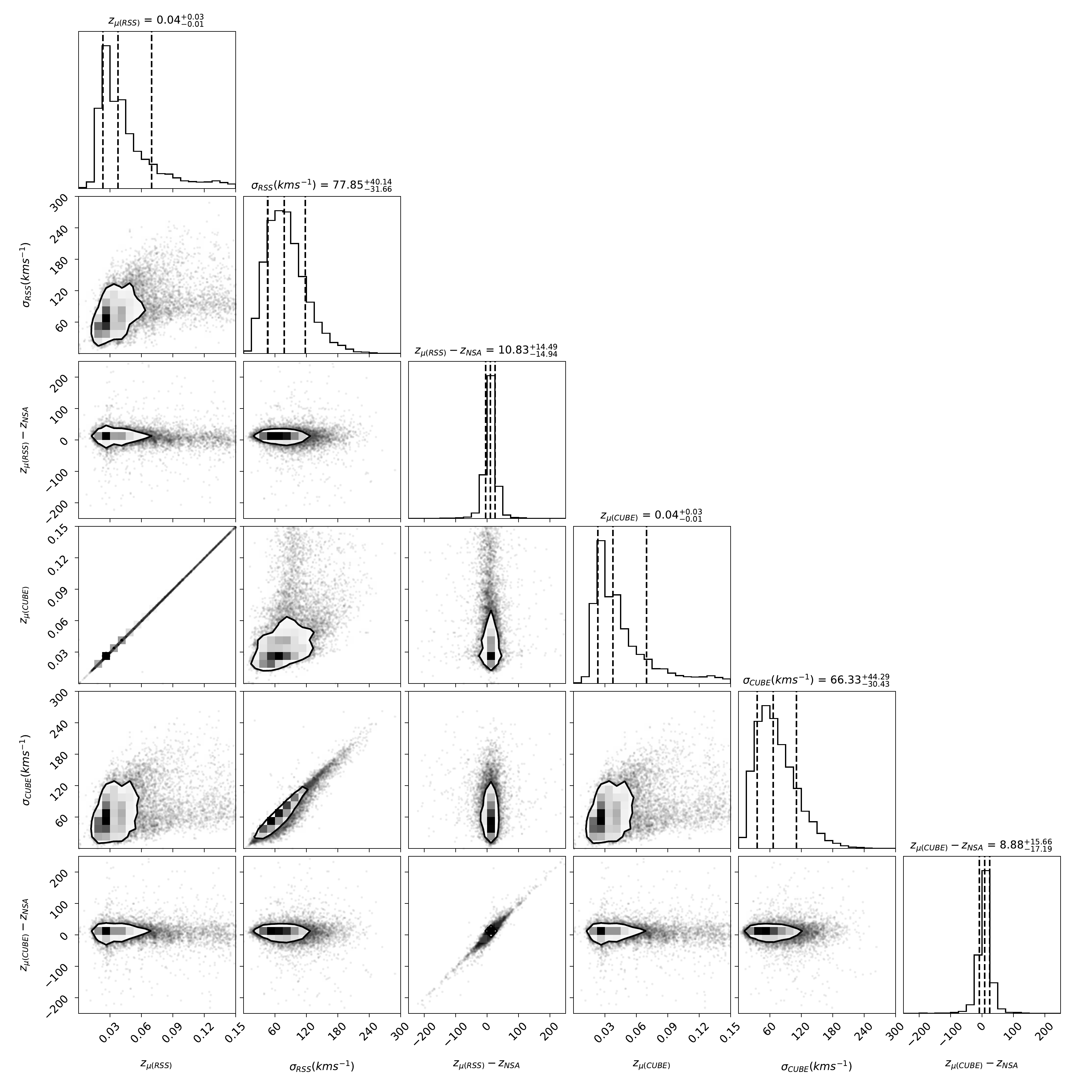}
\end{center}
\caption[Specz Galaxies]{The corner plot demonstrates the redshift properties of galaxies evaluated by the SILO software. Contour boarders in two dimension plots represent the 68\% confident region, with shading within binned regions scaling with the counts per bin. From left to right, the dashed lines within the histograms at the top of each column represent the 16, 50, and 84 quartiles, respectively. No significant bias is detected when comparing the mean of the RSS ($z_{\mu(RSS)}$) spectroscopic redshifts evaluated for each galaxy to the NSA catalogue $z_{\mu(RSS)}-z_{NSA}$, which is also reflected when comparing the mean of the CUBE ($z_{\mu(CUBE)}$) spectroscopic redshifts evaluated for each galaxy to the NSA catalogue $z_{\mu(CUBE)}-z_{NSA}$. The standard deviations of the RSS ($\sigma_{RSS}$) and spaxel ($\sigma_{RSS}$) sample for each galaxy reflect the expected LOS velocity profiles of galaxies evaluated within the field of view of MaNGA IFUs.}
    \label{fig:specz_galaxy_projections}
\end{figure*}

Spectroscopic redshifts are successfully computed for 5,398,665 RSS across 10,204 MaNGA targets and 15,973,915 spaxels across 10,714 MaNGA targets. The distribution of galaxy redshifts evaluated for the sample has a median redshift of 0.037 and an interquartile range (IQR) of 0.028, and is similar to the photometric redshift distribution of the NSA catalogue (see Figure~\ref{fig:specz_galaxy_projections}) since the median difference is $\approx$10 km\,s$^{-1}$ with a scatter of $\approx$15 km\,s$^{-1}$. The median spectroscopic redshift uncertainty ranges from 11 km\,s$^{-1}$ for spectra with a mean $SN\ge3$ to 52 km\,s$^{-1}$ for spectra with a mean $SN<3$, enabling the LOS velocity profile to be resolved for targets (see Figures~\ref{fig:speczexample} and~\ref{fig:specz_galaxy_projections}) while being at least an order of magnitude improvement over SDSS photometric uncertainties~\citep{2005astro.ph..8564S}. The spectroscopic redshifts and four-component PCA fit are presented in the Specz VAC (see Section~\ref{speczvac}) to the public for use in various measurements that require precision modeling and redshift measurements.

\subsection{Candidate Background Emissions}

The SILO software scanned 1,247,568 co-added residuals stacked across 5,402,098 foreground-subtracted RSS from the full MaNGA target sample to find 80,954 multi-line and 33,738 single-line detections, of which 7,158 multi-line and 1,124 single line detections passed the pre-inspection cuts. Out of 8,282 detections of candidate sets of background emission-lines that passed the pre-inspection cuts, manual inspection of the spectra revealed 1,042 are likely (Grade A), 37 are probable (Grade B), and 26 are possibly (Grade C) from background galaxies. The redshift grouping of the detections across 441 targets suggests 458 candidate source planes containing one or more background galaxies. The counts per grade are listed in Table~\ref{table:hits_in_sources}.

\subsection{Strong Lensing Projections}\label{er_results}

\begin{figure*}
\centering
\begin{center}
\centering
\subfloat{
    \includegraphics[height=3.33cm]{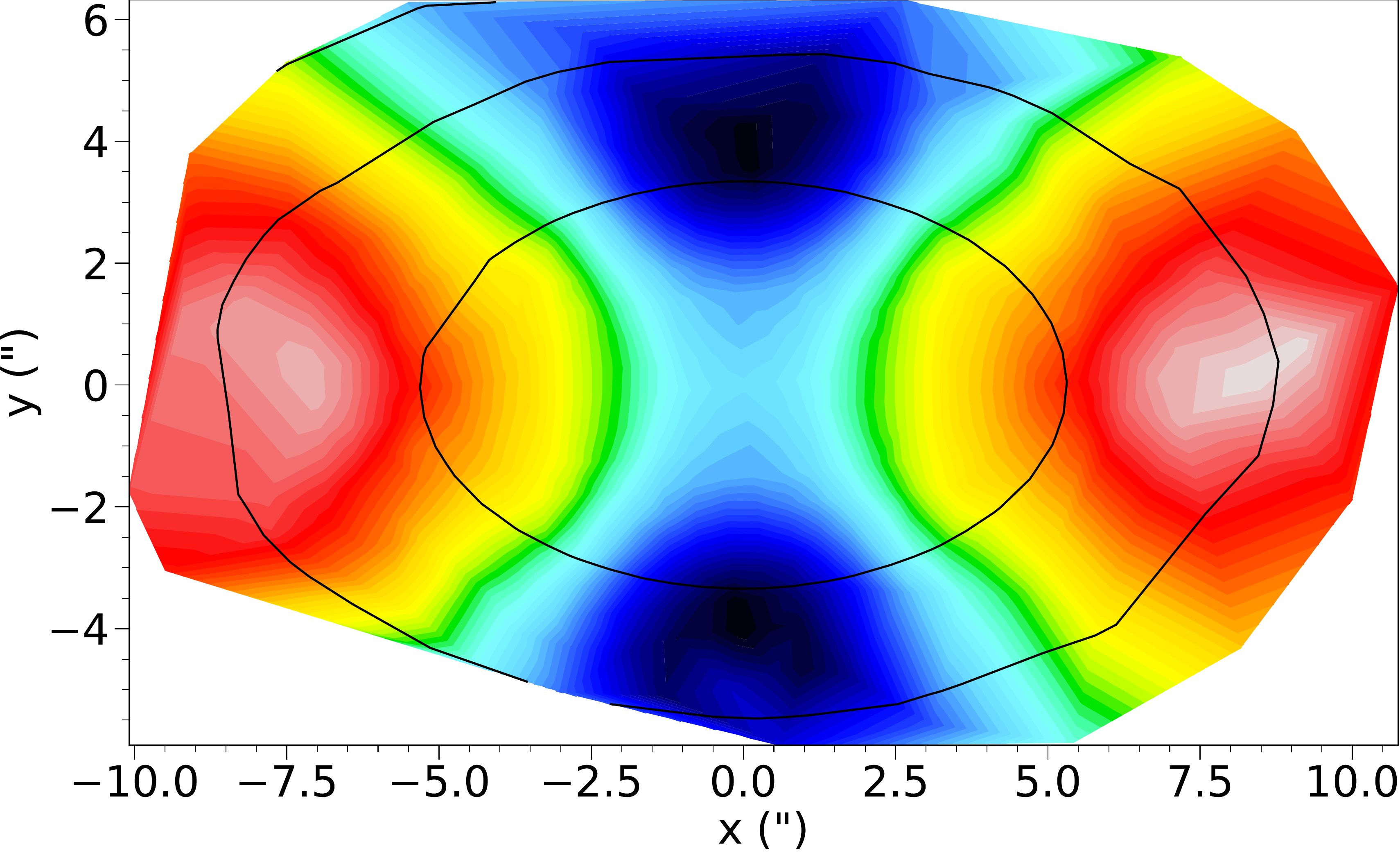}
    \includegraphics[height=3.33cm]{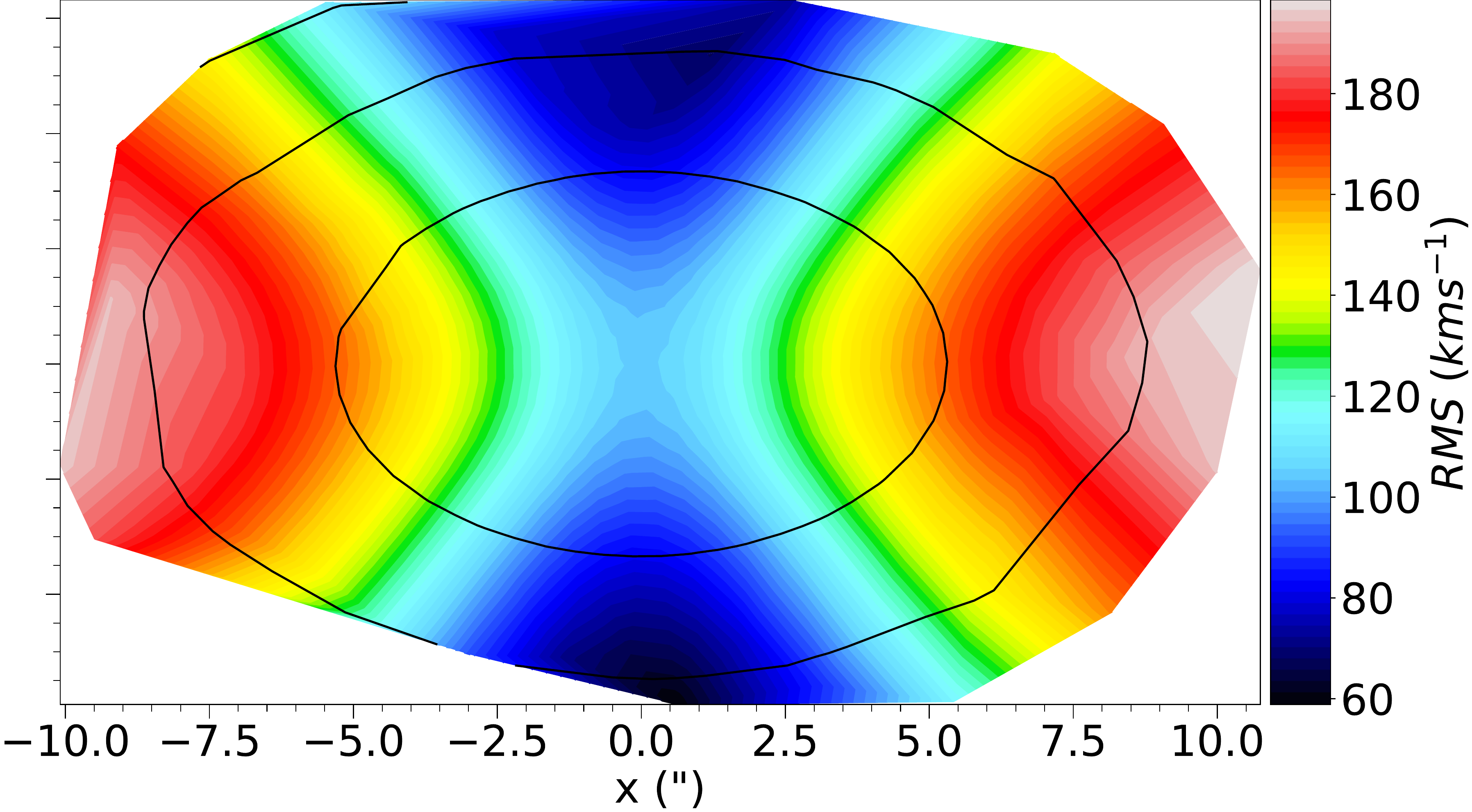}
    \includegraphics[height=3.33cm]{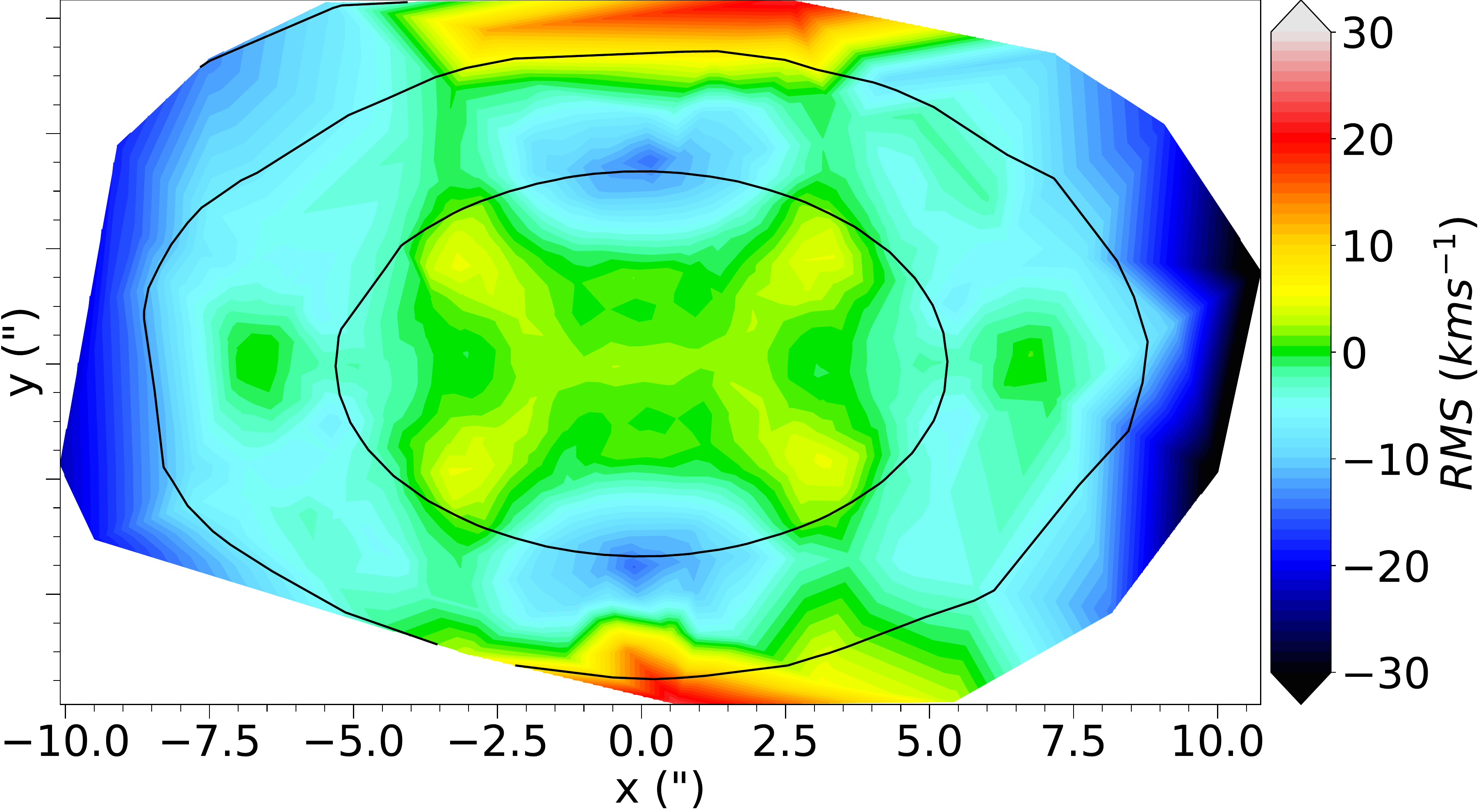}
}
\end{center}
\caption[vel example]{Example of a JAM fit to galaxy kinematics from MaNGA. The color of the top left (data symmetrized by the \texttt{symmetrize\_velfield} function contained within the JAM package), top middle (JAM model), and right (residual) contour plots for system SDSS J1344+2620 represent the scaling in the color bar of the root-mean-square of the combined stellar velocity and velocity dispersion obtained from MaNGA kinematic maps, with a separate color bar for the residual panel. The peaks in the velocity field have been aligned horizontally.}
    \label{fig:velexample}
\end{figure*}

\begin{figure*}
\begin{center}
\includegraphics[width=\textwidth]{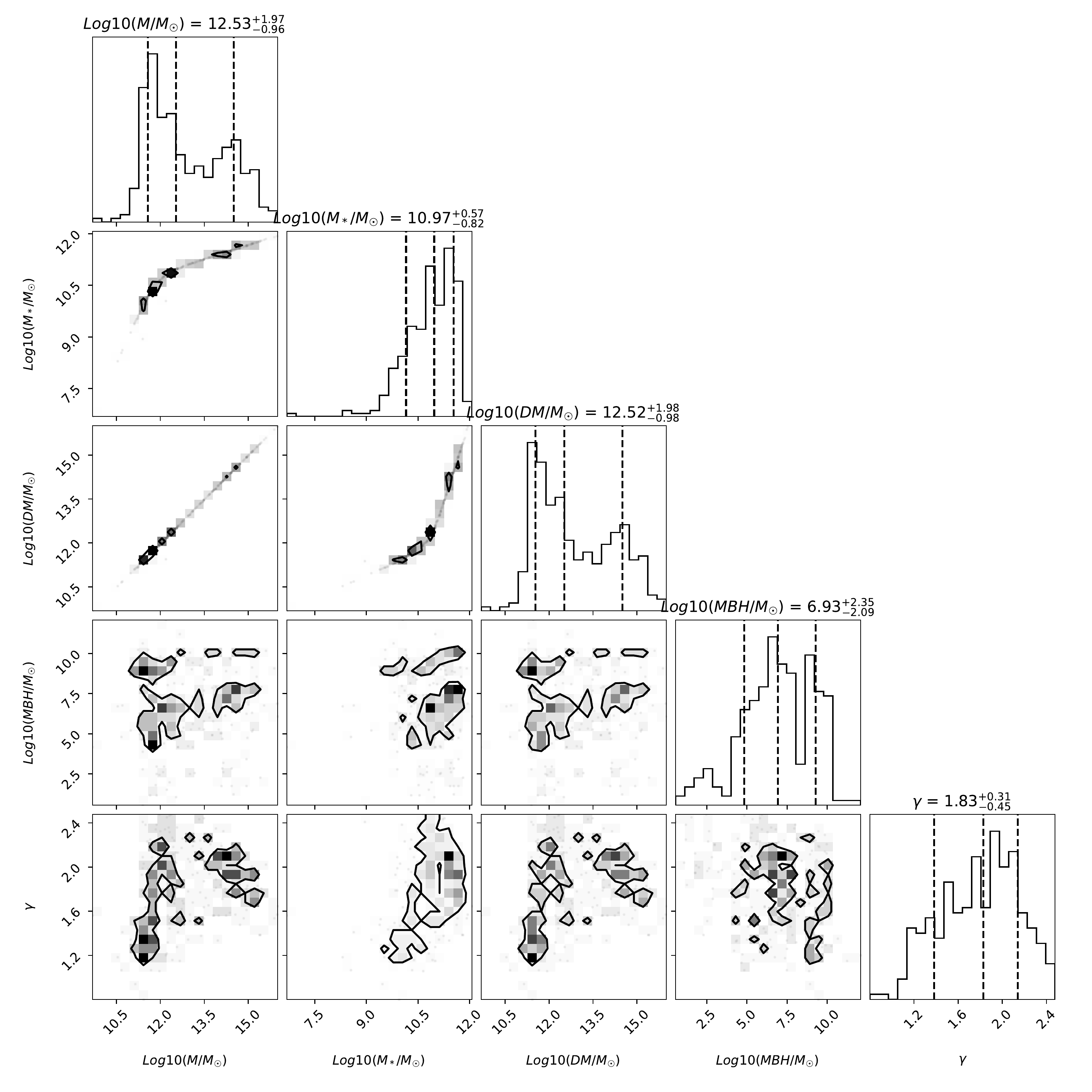}
\end{center}
\caption[JAM mass fits]{Successfully computed JAM mass projections of 302 MaNGA galaxies with at least a detection within eight arcseconds from galaxy center that belongs to an assured background galaxy candidate, which radial cut is explained in Section~\ref{densitycomputation}. Contour boarders in two dimension plots represent the 68\% confident region, with shading within binned regions scaling with the counts per bin. From left to right, the dashed lines within the histograms at the top of each column represent the 16, 50, and 84 quartiles, respectively. The scaling relations between the total mass ($M$), stellar mass ($M_*$), and dark matter ($DM$) reflect literature projections~\cite{2014MNRAS.441.3359D, 2017AJ....154..190H, 2018MNRAS.481.2813G, 2020A&A...634A.135G}. The scaling of the black hole mass ($MBH$) is approximate to as described in literature~\citep{2009ApJ...704.1135B}. The fit of the power-law density slope ($\gamma$) within twice the effective radius  approximates the isothermal slope ($\gamma$=2) observed in massive lenses~\cite{2006ApJ...640..662T}.}
    \label{fig:jam_mass_fits}
\end{figure*}

Figure~\ref{fig:velexample} demonstrates that fitted JAM kinematic projections of MaNGA target SDSS J1344+2620~approximate the DAP kinematic measurements used in the fit. Distribution of successfully computed JAM model properties for 302 examined source-planes listed in Table~\ref{table:sourcecount} are demonstrated in Figure~\ref{fig:jam_mass_fits}. Inspection of kinematic maps reveals JAM fails to fit the remaining quarter of targets with candidate background galaxies. This issue is primarily caused by poorer kinematic measurements of fainter and low-mass galaxies failing to pass masking and fitting processes described in Appendix~\ref{jamdynamics}. Fortunately, the impact of failed fits is negligible to our lens search method since the probability of lensing scales with galaxy mass. In addition, kinematic measurements are not required in the UER computation method described in Sections~\ref{fireflyer} and Appendix ~\ref{apdx_stellar_mass_profile}, which enables at least one UER measurement for 96\% of the examined source planes.

The distribution of the JAM models for the 458 examined source-planes listed in Table~\ref{table:sourcecount} are demonstrated in Figure~\ref{fig:jam_mass_fits}. The scaling between the stellar mass and the dark matter halo properties in Figure~\ref{fig:jam_mass_fits} originate from relations provided by theory and simulations within literature~\citep{2014MNRAS.441.3359D, 2020A&A...634A.135G}. Thus the scaling between the stellar mass and the dark matter halo is fixed in the JAM modeling process and demonstrate no scatter. This stellar to dark matter scaling is desirable since MaNGA kinematic measurements are only well constrained within approximately an effective radius ($R_{eff}$) of the target, which is insufficient for a fit to isolate the low dark matter fraction within this region. Thus the scaling of the total mass to the dark and stellar mass is also fixed and shows no scatter in Figure~\ref{fig:jam_mass_fits}.

The scaling and concentration of the theoretical dark matter density profile with an MGE fitted projection of the stellar mass yield several properties similar to real galaxies samples. In particular, the slope of the total density profile, ($\gamma$), depends on the contributions of the dark matter density profile, which scales with an approximate power-law slope of one within the scale radius~\cite{1996ApJ...462..563N}, and the stellar density profile, which is greater than isothermal ($\gamma > 2$). Thus only a strict combination of stellar and dark matter components yield the isothermal slope measured in massive lenses~\citep{2006ApJ...640..662T, 2012ApJ...757...82B, 2018MNRAS.480..431L}. A power-law density slope $\gamma$ is fitted to each total density MGE within $2\,R_{eff}$ and displayed in Figure~\ref{fig:jam_mass_fits}, which reveals an approximately isothermal slope for massive MaNGA galaxies. The median and 86\% confidence levels of the LOS DMF is $0.18^{+0.21}_{-0.11}$ for $R/R_{eff}=0.1$ and $0.48^{+0.19}_{-0.19}$ for $R/R_{eff}=1$, which is consistent with DMF measurements from microlensing~\cite{2015ApJ...799..149J}. The scaling of $\gamma$ with mass is consistent with the results of~\cite{2019MNRAS.490.2124L}, who also used JAM to fit MaNGA galaxies.

\begin{figure*}
\begin{center}
\includegraphics[width=\textwidth]{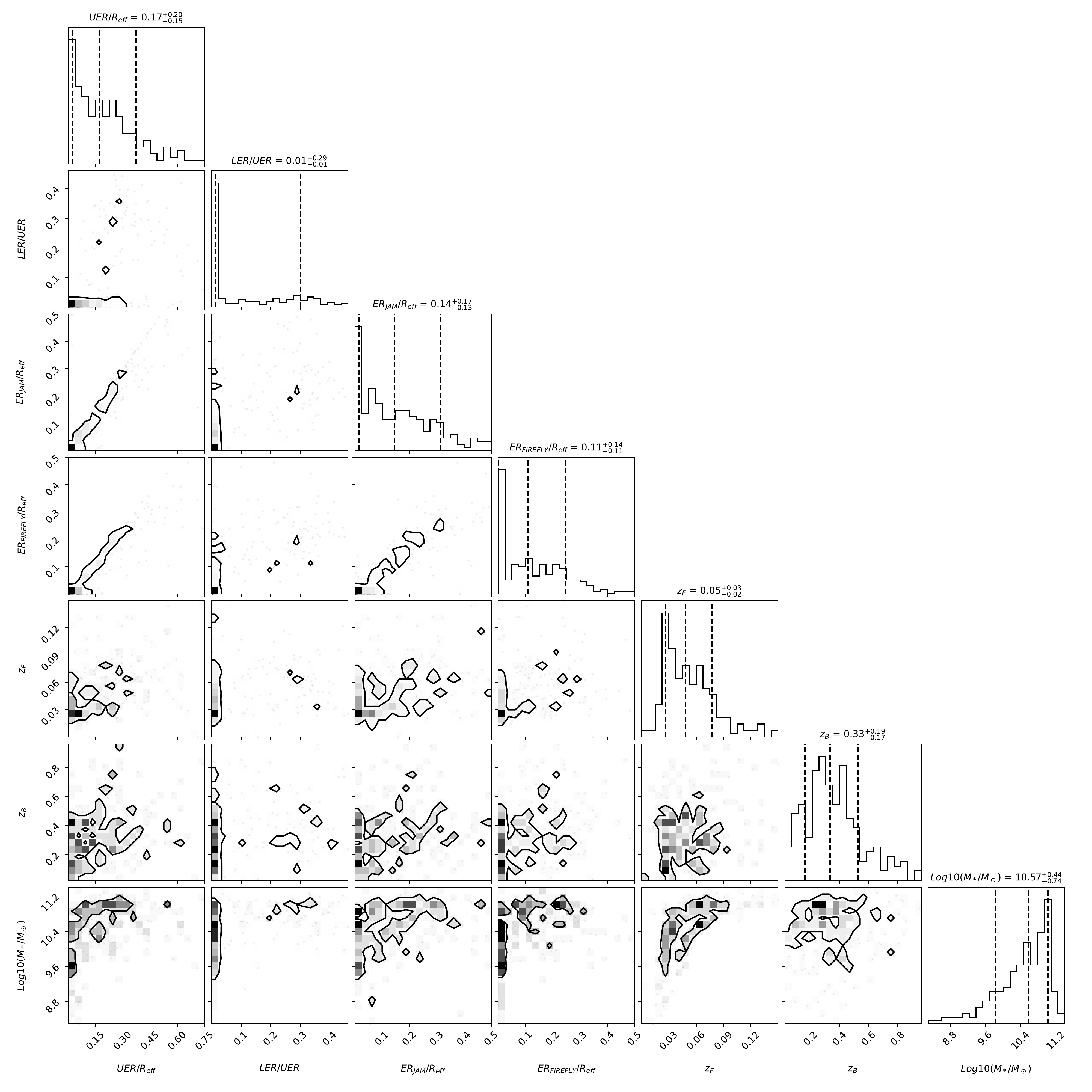}
\end{center}
\caption[JAM er fits]{Einstein radii projection comparisons for 212 candidate source-planes with an $UER>0\arcsec$ and $UER<2ER$ from kinematic-based projections, which the latter filter is set to remove a low fraction of largely uncertain fits. Contour boarders in two dimension plots represent the 68\% confident region, with shading within binned regions scaling with the counts per bin. From left to right, the dashed lines within the histograms at the top of each column represent the 16, 50, and 84 quartiles, respectively. The upper limit of the Einstein radius ($UER$), the Einstein radii projection from JAM models ($ER_{JAM}$), and the Einstein radii projection from stellar-mass maps ($ER_{FIREFLY}$) across the Petrosian fitted effective radius $R_{eff}$ demonstrates a weak scaling with the foreground redshift ($z_F$) and the stellar-mass ($M*$) of the targets, with too much scatter to validate a scaling with the background redshift ($z_B$). The $ER_{FIREFLY}/ER_{JAM}$ is approximately one. The demonstrated stellar mass is the total stellar mass computed from the NSA catalogue, which is a K-correction fit to a fitted elliptical Petrosian model. The ratio between the lower limit of the Einstein radii projections (LER) and the upper limit tends to increase with mass.}
    \label{fig:jam_er_projections}
\end{figure*}

Figure~\ref{fig:jam_er_projections} compares distributions of 219 UER measurements from both computation methods where at least one of the UER measurements is greater than zero arcseconds, and the JAM-based UER measurement is less than twice the JAM-based Einstein radius measurement. The latter filter for Figure~\ref{fig:jam_er_projections} removes 50 JAM-based UER computations that inspection revealed are beyond the typical Einstein radii computations due to inflated uncertainties when the JAM fitting process resolves two or more local $\chi^2$ minima regions. Figure~\ref{fig:jam_er_projections} demonstrates the remaining JAM and FIREFLY-based UER measurements are similar in radius, significantly scale with mass, and more weakly scale with distances to lens and source. The robustness of Einstein radii projections is roughly inferred by the ratio between the lowest and largest Einstein radii ratio demonstrated in Figure~\ref{fig:jam_er_projections}, which reveals the robustness increases with galaxy mass.

\subsection{Narrow-band Comparisons}

\begin{figure*}
\begin{center}
\subfloat[SDSS~J1701+3722]{\includegraphics[width=\textwidth]{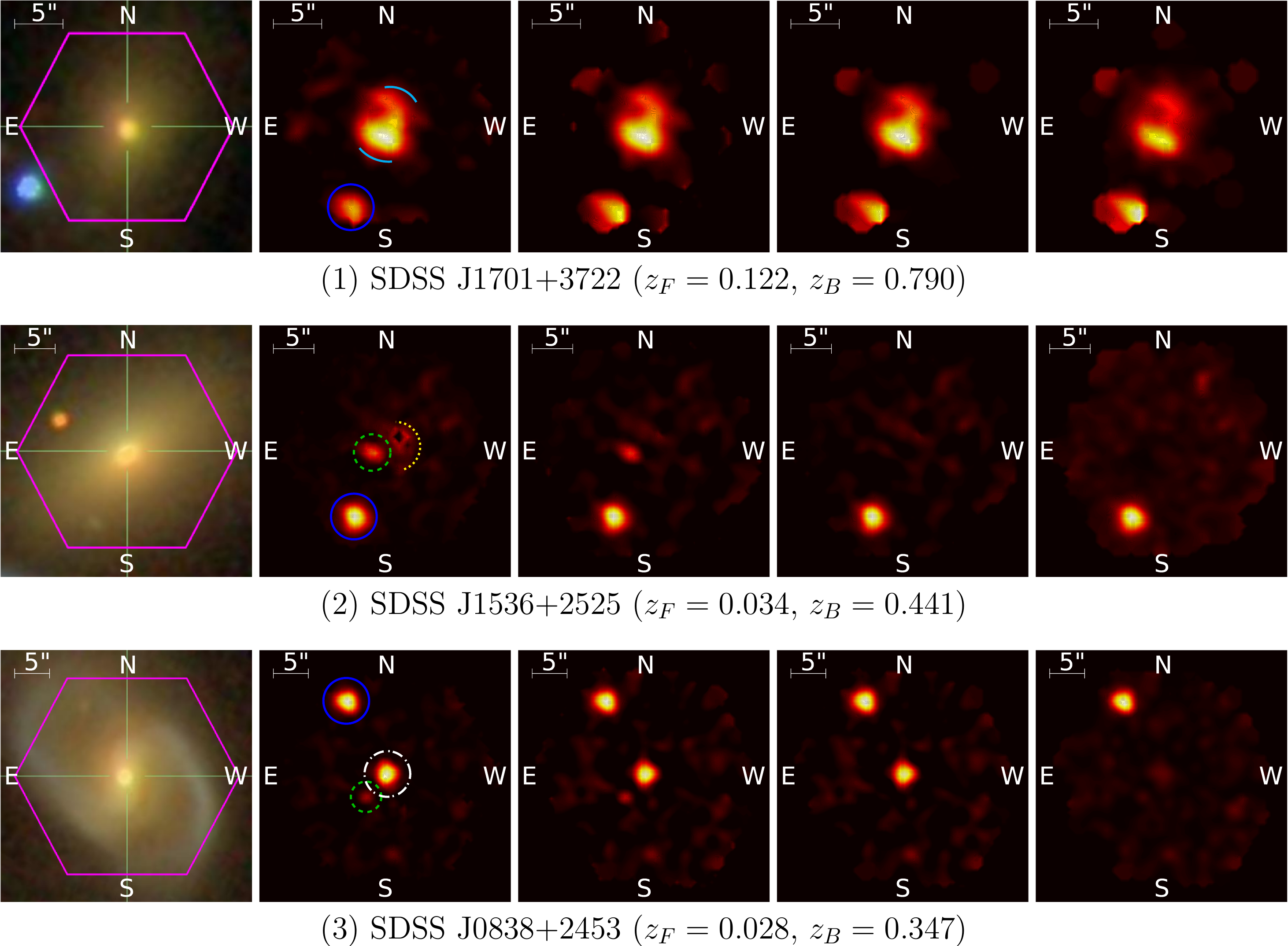}}
\end{center}
\caption[Narrowband Comparisons]{Demonstration of how transient and reduction contamination is suppressed in SN narrow-band images constructed from co-added foreground-subtracted residuals. From left to right, the plots display; an SDSS image of the target galaxy, a narrow-band image constructed from subtracting the flux from MaNGA spaxels, a narrow-band image created from SILO spaxels constructed from foreground subtracted residuals, a narrow-band image created from SILO spaxels constructed from co-added residuals, and an SN narrow-band image created from spaxels of the co-added SN. The shape of unlikely lensed (highlighted by blue circles) and likely lensed (highlighted by cyan arcs) background galaxies varies little across the different types of narrow-band images. The weak signal suppressed near the target center within the SN narrow-band image may either be a lensing feature or contamination from the foreground galaxy bulge. In contrast, improperly subtracted cosmic-rays (highlighted by green dashed circles) do not appear within narrow-band images constructed from co-added residuals since sigma-clipping rejects signals not present across stacked exposures. The dotted yellow arc highlights a flux bias induced by masking the cosmic-ray within nearby RSS, forcing the MaNGA spaxel construction method to stack the relatively higher fluxes within the remaining nearby RSS. The flux bias is not present in the other narrow-band images since the residuals used to construct these spaxels have a mean flux of zero. The white dash-dot circle highlights contamination induced by improper subtraction of lower quality RSS, in which contamination is suppressed within the SN narrow-band image since uncertainty measurements are larger within the affected region.}
    \label{fig:narrowbandcompare}
\end{figure*}

Four types of narrow-band images are inspected across 458 candidate source planes for signs of transient and reduction contamination described in Section~\ref{ch4narrowband} and demonstrated in Figure~\ref{fig:narrowbandcompare} to test if SN narrow-band images are the most ideal for use in source-plane inspection. Central flux bias (see feature within yellow dotted arc in Figure~\ref{fig:narrowbandcompare}) is detected within $16\%$ of narrow-band images created from MaNGA spaxels in which the foreground was subtracted by the four-component PCA model included in the Specz VAC (second column from left in Figure~\ref{fig:narrowbandcompare}), which contamination is not present in narrow-band images created from spaxels constructed from RSS that the foreground was subtracted from a seven-component PCA model (middle column in Figure~\ref{fig:narrowbandcompare}) since the mean of the residuals is zero during spaxel construction. Since flux bias resembles arcs or counter images in $2\%$ of narrow-band images constructed from foreground subtracted MaNGA spaxels, we validate that spaxels constructed from residuals are relatively cleaner of contaminants.

Incompletely masked transient contamination (see features within dashed green circles of Figure~\ref{fig:narrowbandcompare}) is clearly visible in $3\%$ of narrow-band images constructed either from foreground-subtracted MaNGA spaxels or spaxels constructed from individual residuals, which features are not present in narrow-band images created from spaxels constructed from co-added residuals (third column from left in Figure~\ref{fig:narrowbandcompare}) since transient signals are not present across most exposures and thus are rejected in the sigma-clipping process. Since this contamination randomly occurs across the entire narrow-band image, only $5\%$ resembles a near centre counter-image, which translates to $0.2\%$ could be miss-identified as strong lensing features. Thus we validate that spaxels constructed from co-added residuals are relatively cleaner than either of the previously compared narrow-band types.

Non-perfectly subtracted features (see features within white dash-dot circles of Figure~\ref{fig:narrowbandcompare}) are present within $11\%$ of narrow-band images constructed by integrating over flux of any form of residuals, which features are suppressed in narrow-band images created from spaxels that are constructed from the SN of co-added residuals (right column in Figure~\ref{fig:narrowbandcompare}) since poorer model fits are often related to spectra with uncertainties inflated by reduction issues. The features often appear near the galaxy centre since residuals can scale with the difference of the brighter flux and the foreground model. Inspection revealed $18\%$ of these central features are positioned where a counter-image would be expected relative to the candidate background galaxy, which translates to $2\%$ of narrow-band images created from integrating over flux of any form of residuals shows candidate counter-images that cannot be assured by imaging alone. The signal in the spectra at the location of these low-SN candidate counter images is often inconclusive or suggestive of contamination.

Unfortunately, the SN narrow-band images may suppress real counter images whose signal is below the noise. The shape of unlikely lensed galaxies (see features within blue circles of Figure~\ref{fig:narrowbandcompare}) and galaxies with visible strong lensing features (see features within cyan arcs of Figure~\ref{fig:narrowbandcompare}) is nearly identical across narrow-band image types while faint possible counter images are suppressed in $4\%$ of the SN narrow-band images. The signal in the spectra at the location of the suppressed candidate counter images is often inconclusive or suggestive of contamination, and thus we suspect only a fraction of these low-SN counter images are real. However, possible candidate loss from SN suppression is mitigated in SILO source-plane inspection since a candidate source is also graded on its proximity to the probable strong-lensing regime. Thus we concluded SN narrow-band images are ideal to use in the SILO source-plane inspection method.

\subsection{Candidate Sources}

Spectra inspection revealed $98\%$ of 458 candidate background galaxies across 441 targets contained one or more highly assured detections. The source-plane inspection revealed 8 likely, 17 probable, and 69 possible candidate sources, with counts per source grade listed in Table~\ref{table:sourcecount}.

Figure~\ref{fig:sourcesexample} displays the spatial position(s) of detections is at or within the largest UER for source-planes with a grade of A- or above. In addition, Figure~\ref{fig:sourcesexample} also shows one or more features supportive of lensing within SN narrow-band images of the source-planes. Relative to the brightest candidate source image, a fainter candidate counter-image is located nearer to and on the opposite side of the candidate lens for SDSS J1701+3722, SDSS J1712+3014, SDSS J1509+3015, and SDSS J1341+5538, which multiple source images are required to define a system as strongly lensed. System SDSS J1701+3722 was previously detected by~\cite{2017MNRAS.464L..46S} and later by the first SILO scan of MaNGA~\citep{2018MNRAS.477..195T}, which follow-up with improved IFU observations support the most northern and southern bright features within the SN narrow-band image are from two different background galaxies while the brightest image and the small faint image located between the two bright central images are from the same source~\citep{2020MNRAS.493L..33S}. Since SN narrow-band images can suppress faint counter images near the target center, we do not demote the remaining half (SDSS J1308+3400, SDSS J1632+1338, SDSS J0754+3910) of candidates with A- or above grades, especially since these systems demonstrate features supportive of arcs observed in strong lensing of extended sources.

The source grade of assurance strongly scales with galaxy mass since the computation of the probable strong lensing regime and the separation between lensed source images observed within SN narrow-band images scale with the square of the enclosed mass. The scaling is evident in Figure~\ref{fig:sourcesexample} since all candidate lenses with a grade of A- or above are early-type massive galaxies. To quantify this trend further, we compared lens candidate grades with a stellar-mass approximation obtained from the NSA catalogue, in which the stellar mass is obtained by a K-correction fit of an elliptical Petrosian model to SDSS photometry that is then multiplied by an estimate of the mass-to-light ratio. The comparison revealed that lens candidates with $\ge10^{11}$ solar masses are present in all grades. The lowest mass is $\ge10^{10.48}$ solar masses for grades above A-, $\ge10^{9.73}$ solar masses for grades above B-, and $\ge10^{8.3}$ solar masses for grades above C-. Thus follow-up high-resolution imaging of probable and more highly assured lens candidates should yield on the order of ten low-redshift massive lenses to test if the lens-dynamic mass discrepancy is present when comparing the lens mass with dynamic mass measurements from follow-up improved IFU kinematics, which validation of the discrepancy will warrant further investigation if LOS mass, the dark matter halo, or the cluster halo adds enclosed mass to lens measurements. Confirmation of a set of lower-mass lenses within the probable and possible candidates can be combined with massive lenses to determine if the discrepancy scales with lens mass and is thus more likely caused by either the galaxy or cluster halo, which the latter can be related to the visible environment around the lens. The lens and improved dynamic measurements of a sample of confirmed young low-mass lenses can also be compared with the 20 ELG lenses found by the Sloan WFC Edge-on Late-type Lens Survey~\citep[SWELLS;][]{2011MNRAS.417.1601T, 2011MNRAS.417.1621D} to potentially break the degeneracy between the bulge, disk, and halo mass components while measuring changes in the density profile between low-mass and high-mass lenses at cosmological distances.

The lens candidates are distributed between redshifts of 0.02 and 0.15, and thus improved joint lens-IFU based dynamic measurements can tightly constrain the density profile of confirmed massive lenses within MaNGA for comparison to massive lenses found in other SDSS surveys at higher redshifts (0.1 < z < 0.7) including the Sloan Lens ACS ~\citep[SLACS;][]{2006ApJ...638..703B, 2015ApJ...803...71S, 2017ApJ...851...48S} survey and BELLS~\citep{2012ApJ...744...41B, 2012ApJ...757...82B} program. Tight profile constraints from the improved joint lens-dynamic measurements will reduce the number of low-redshift massive lenses required to statistically improve measurements of the inferred evolution in the power-law density slope found in higher redshift lenses.

Projections of the Einstein radii from mass models are of the order of an arcsecond or less, which translates into an order of half an effective radius or less for MaNGA targets. Thus a sample of confirmed massive lenses within MaNGA can be used to probe the baryon-dominated bulge of massive galaxies, which previous SDSS lens surveys have not constrained since higher-redshift LRG lens samples found within SDSS contain physical probe radii typically larger than half an effective radius~\citep[see Figure 4 of][]{2012ApJ...744...41B}. In addition, joint lens-dynamic measurements of the density profile within the inner region of MaNGA lenses can be combined with other SDSS lens samples to extend radial evolution inferences of the density profile~\citep{2018MNRAS.480..431L} to the galaxy center. Finally, SDSS J1701+3722 has been used to test which initial mass functions reasonably project the enclosed stellar mass for massive galaxies~\citep{2020MNRAS.493L..33S}, which a sample of low-redshift lenses may be of use to bolster the results statistically.

\newcommand{\sourcescaption}{The location of each detection contained within likely (Grade A- or above) strong galaxy-galaxy gravitational lens candidates. The left image of each set of the target displays the candidate lens galaxy, the field-of-view of the IFU (pink hexagon), the maximum upper limit of the strong lensing regime (lime circle), and the number of candidate source detections within a radius of 0.5" (black marker with a cyan number).  The right image of each set of the target displays a narrowband image created from the total SN of the background emission lines mentioned in the bottom-left corner of the figure. The maximum upper limit of the strong lensing regime is shown as a lime circle. Each image is retrieved from the lens VAC, which is described in Section~\ref{lensvac}. The star by the name indicates if the largest displayed strong lensing regime is computed using the \texttt{FIREFLY} stellar mass maps multiplied by an NFW DM fraction (see Section~\ref{fireflyer}), where no star indicates the largest displayed strong lensing regime is computed by fitting DAP kinematics with JAM (see Section~\ref{cjam}).}

\begin{figure*}
    \centering
    \includegraphics[scale=1.3]{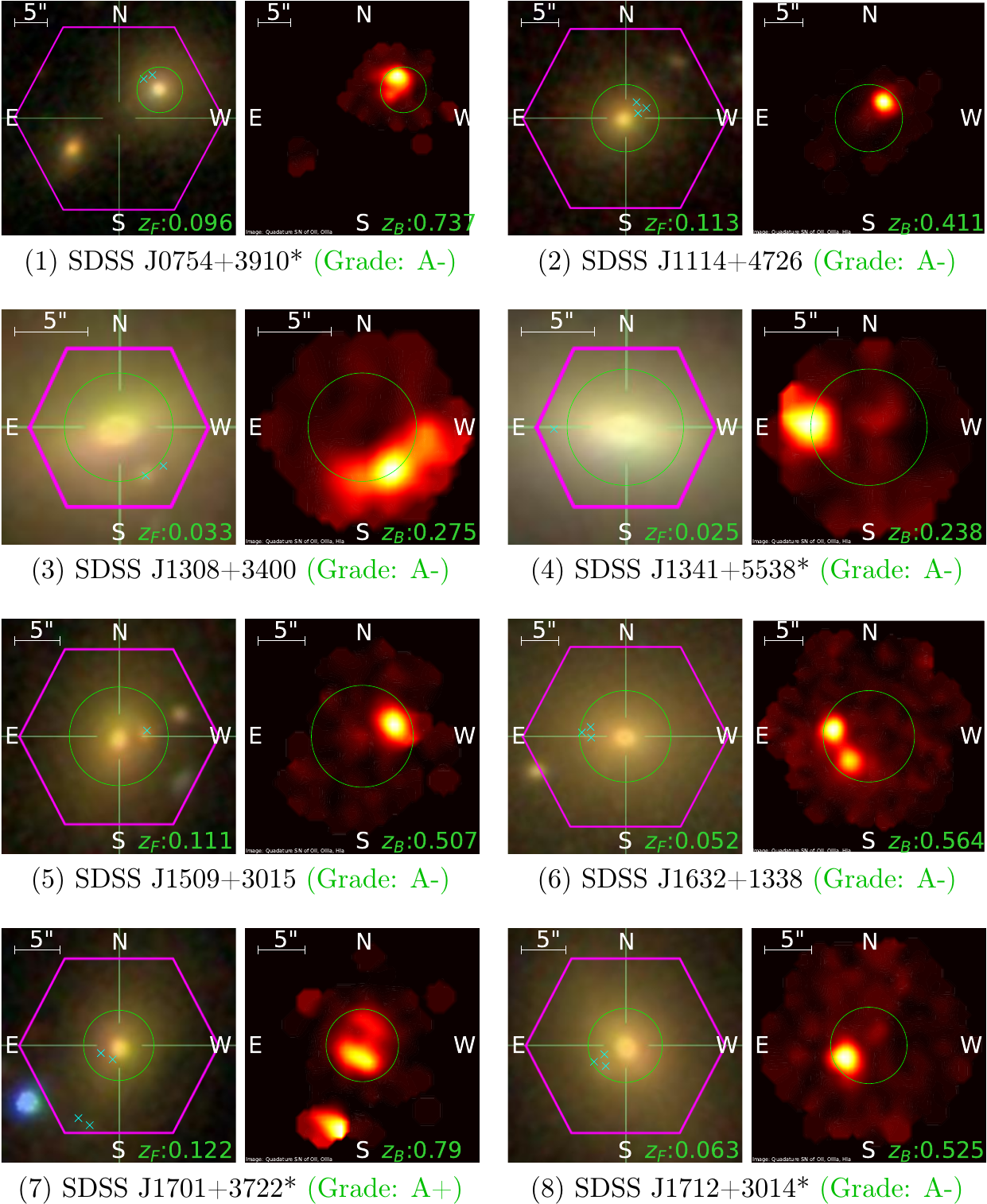}
    \caption[Geometry of Grade-A detections]{\sourcescaption}
    \label{fig:sourcesexample}
\end{figure*}

\section{Presenting the Spectroscopic Redshifts Value-Added-Catalogue}\label{speczvac}

The spectroscopic redshifts computed in Section~\ref{speczcomp} are presented in a Specz VAC scheduled for release in SDSS-IV DR17. The Specz VAC contains 7,910 RSS and 7,992 spaxel reductions that are not flagged as problematic by the DRP or SILO. The Specz VAC also includes a PCA fit of the foreground flux using a four-component Eigenspectra template. The best-fit was applied in Section~\ref{speczcomp} to compute spectroscopic redshifts but not in foreground subtraction.

The Specz VAC is located on the Science Archive Server (SAS)\footnote{Specz VAC: \url{https://data.sdss.org/sas/dr17/manga/spectro/specz/v3_1_1/1.0.1}} of SDSS and is divided into three types of files:
\begin{enumerate}
    \item The Specz VAC contains a summary fits file (\texttt{speczall.fits}) that presents the analysis information, MaNGA target information, and statistics on the evaluated spectroscopic redshift samples for each target. The extensions of the fits file are separated into survey and target information in the primary extension, as shown in Table~\ref{table:speczall_primary}, and statistics on the spectroscopic redshift samples of the target in the \texttt{SUMMARY} extension, as shown in Table~\ref{table:speczall_summary}. The columns of this fits file are described and documented in the data model located on the SDSS server.\footnote{Specz VAC summary model: \url{https://data.sdss.org/datamodel/files/MANGA_SPECZ/DRPVER/SPECZVER/speczall.html}}
    \item The Specz VAC contains a fits file (\texttt{specz-PLATE-IFU-LOG[FILETYPE].fits}) for each presented reduction. The fits file contains the spectroscopic redshifts and the foreground flux models for each successfully computed spectroscopic redshift within three $\sigma_{MLE}$ of $\mu_{inner}$. The primary extension of the fits file contains galaxy information, the mean and standard deviation of the redshifts within $R_{inner}$, and the galaxy mean and sigma of the spectroscopic redshifts contained within the fits file, as shown in Table~\ref{table:specz_vac_table_primary}. The row index of the RSS (and column index for spaxels), spectroscopic redshift and uncertainties, and spatial position of the spectra are in the \texttt{REDSHIFTS} extension, as shown in Table~\ref{table:specz_vac_table_redshifts}. The foreground models are in units of \fluxunits~and are presented in the \texttt{MODEL} extension. The columns of this fits file are described and documented in the data model located on the SDSS server.\footnote{Spectroscopic redshifts VAC data model: \url{https://data.sdss.org/datamodel/files/MANGA_SPECZ/DRPVER/SPECZVER/PLATE4}}
    \item The Specz VAC contains png and pdf plots of both the spectroscopic redshifts and the mean SN of the spectra relative to the galaxy centre (see Figure~\ref{fig:speczexample} for an example). These plots are located in the \texttt{images} folder within each plate folder, with paths of the form \texttt{PLATE/images/specz-PLATE-IFU-LOG[FILETYPE].EXTENSION}.
\end{enumerate}

\section{Presenting the Lens Candidates in a Value-Added-Catalogue}\label{lensvac}

As in~\cite{2020arXiv200709006T}, presenting the candidates to the public in fits files enables more data to be available for various research interests within the lensing community. The candidates are presented in a lens VAC being released in DR17. The data released will contain information on each source or background galaxy candidate as well as processed spectra that is scanned for background emission-lines. The lens VAC is located on the SAS\footnote{Lensing VAC: \url{https://data.sdss.org/sas/dr17/manga/spectro/lensing/silo/v3_1_1/1.0.4}} and is divided into three types of files:
\begin{enumerate}
    \item The lens VAC contains a summary fits file (\texttt{silo\_manga\_detections-SILOVERSION.fits}) that presents spatial and spectroscopic data on each background galaxy candidate. The extensions of the fits file are separated into survey information and emission-lines searched in the primary extension, as shown in Table~\ref{table:vac_table_overview}, candidate lens source-plane statistics and inspection information, as shown in Table~\ref{table:vac_table_source}, detection and spectra inspection information, as shown in Table~\ref{table:vac_table_detection}, and candidate emission-line analysis, as shown in Table~\ref{table:vac_table_analysis}. The Gaussian fit and analysis is recorded for any candidate emission-line with \(SN > 3\). The columns of this fits file are described and documented in the data model located on the SDSS server.\footnote{SILO MaNGA summary VAC data model: \url{https://data.sdss.org/datamodel/files/MANGA_SPECTRO_LENSING/silo/DRPVER/SILO_VER/silo_manga_detections.html}}
    \item The lens VAC contains a fits file for each candidate background galaxy (\texttt{manga\_PLATE\_IFU\_stack\_data.fits}) that includes the foreground-subtracted RSS, the stacked residuals, and the SILO constructed spaxels. The primary extension contains target information, as shown in Table~\ref{table:vac_table_primary_stack}. The \texttt{INDIVIDUAL\_EXPOSURE\_REDUCTION} extension contains the RSS data, residuals, SN of the residuals, and the position information, as shown in Table~\ref{table:vac_table_rss}. The \texttt{STACKED\_REDUCTION} extension contains the co-added spectra, the SN of the stack, and the positions of the stack, as shown in Table~\ref{table:vac_table_stack}. The extensions containing the MaNGA spectra resolution data and the SILO constructed spaxels of the flux, residuals, and SN are shown in Table~\ref{table:vac_table_spaxels}. The columns of this fits file are described and documented in the data model located on the SDSS server.\footnote{SILO MaNGA spectra data model: \url{https://data.sdss.org/datamodel/files/MANGA_SPECTRO_LENSING/silo/DRPVER/SILO_VER/PLATE/manga_PLATE_IFU_stack_data.html}}
    \item The lens VAC contains pdf plots of the candidate emission-lines with a \(SN > 4\), which plots are located in the \texttt{images} folder and are organized by plate, with paths in the form of \texttt{images/PLATE/flux-PLATE-IFU-FIBERPOSITION-DETECTIONID.pdf}. The inspection grade and comments are displayed below each plot, which comments are not limited to emission-lines with a \(SN > 4\) since low SN emission-lines may also be observed during the spectra inspection.
    \item The lens VAC contains low-resolution images of the target galaxy from SDSS that are overlayed with the locations of the detections and the maximum UER, as shown in Figure~\ref{fig:sourcesexample}. These files are organized by plate within the \texttt{images} folder, with paths of the form \texttt{images/PLATE/silo\_manga\_image-PLATE-IFU-BACKGROUNDID.pdf}.
    \item The VAC contains SN narrow-band images of the candidate source-plane overlayed with the maximum UER, as shown in Figure~\ref{fig:sourcesexample}. These files are organized by plate within the \texttt{images} folder, with paths of the form \texttt{images/PLATE/silo\_manga\_narrowband\_image-PLATE-IFU-BACKGROUNDID.pdf}.
\end{enumerate}

\section{Conclusions}\label{ch4conclusions}

The SILO project has completed its spectroscopic scan of the completed MaNGA survey to find 8 likely, 17 probable, and 69 possible candidate gravitational lenses. The search method improves upon the spectroscopic selection method applied in~\cite{2018MNRAS.477..195T} and~\cite{2020arXiv200709006T} to scan co-added residuals for candidate background detections, use SN narrow-band images to observe any source features, and compare the proximity of each candidate background galaxy to the probable strong lensing regime.

SILO scanned 1,247,568 co-added residuals across 5,402,098 RSS from the MaNGA reductions to reveal 114,692 detections, of which 8,282 detections passed the pre-inspection cuts, which cuts are designed to reject sky and target emission-line contamination. Spectra inspection revealed 1,043 likely, 37 probable, and 26 possible background detections. Source-plane inspection revealed 8 likely, 17 probable, and 69 possible lensed sources across 441 targets with one or more candidate background galaxies.

Whether these detections are quantitatively consistent with a spectroscopic lensing probabability function is the subject of a future manuscript that would generalize the single-fiber approach taken in \cite{2012ApJ...753....4A}, which was based on a monte carlo simulation of $o(10^6)$ mock lenses, which was searched for detectable signatures of strong lensing from background emission lines.  The addition of multiple fibers in which the Einstein Radius provides a possible geometric distribution of detectable lensing features would make an update of this simulation an important tool in understanding the completeness of our catalog.

The spectroscopic redshifts computed for the MaNGA RSS and spaxel data are scheduled for release in a (Specz) VAC for DR17. The lens candidates are also scheduled for release in a lens VAC for DR17, which VAC will contain a summary file of the candidates, spectra data files, plots of high SN candidate background emission-lines, low-resolution images overlayed with the detections, and SN narrow-band images of the background or source candidates.

Follow-up high-resolution imaging of these candidate lenses will provide an opportunity to compare previously found SDSS lenses with confirmed MaNGA lenses, which combined sample will be used to test the evolution in the galactic mass profile across a broadened sample range of redshifts. The SILO project can also use MaNGA lenses to test the inner density profile of the lens since the Einstein radii should be located near the baryon-dominated bulge~\citep[see][]{2018MNRAS.477..195T}. The angular size of MaNGA galaxies enables spatially resolved kinematics to be obtained to constrain the dynamic mass better. The SILO project intends to use improved measurements of the dynamic mass to test if the lens-dynamic mass discrepancy is caused by oversimplified modelling or LOS mass contamination. Resolving the lens-dynamic mass discrepancy may reveal improvements to the lens and dynamic modelling methods, so uncertainties in $H_0$ and the evolution of the galactic mass profile can be mitigated.

\section*{Acknowledgements}

Funding for the Sloan Digital Sky 
Survey IV has been provided by the 
Alfred P. Sloan Foundation, the U.S. 
Department of Energy Office of 
Science, and the Participating 
Institutions. 

SDSS-IV acknowledges support and 
resources from the Center for High 
Performance Computing  at the 
University of Utah. The SDSS 
website is www.sdss.org.

SDSS-IV is managed by the 
Astrophysical Research Consortium 
for the Participating Institutions 
of the SDSS Collaboration including 
the Brazilian Participation Group, 
the Carnegie Institution for Science, 
Carnegie Mellon University, Center for 
Astrophysics | Harvard \& 
Smithsonian, the Chilean Participation 
Group, the French Participation Group, 
Instituto de Astrof\'isica de 
Canarias, The Johns Hopkins 
University, Kavli Institute for the 
Physics and Mathematics of the 
Universe (IPMU) / University of 
Tokyo, the Korean Participation Group, 
Lawrence Berkeley National Laboratory, 
Leibniz Institut f\"ur Astrophysik 
Potsdam (AIP),  Max-Planck-Institut 
f\"ur Astronomie (MPIA Heidelberg), 
Max-Planck-Institut f\"ur 
Astrophysik (MPA Garching), 
Max-Planck-Institut f\"ur 
Extraterrestrische Physik (MPE), 
National Astronomical Observatories of 
China, New Mexico State University, 
New York University, University of 
Notre Dame, Observat\'ario 
Nacional / MCTI, The Ohio State 
University, Pennsylvania State 
University, Shanghai 
Astronomical Observatory, United 
Kingdom Participation Group, 
Universidad Nacional Aut\'onoma 
de M\'exico, University of Arizona, 
University of Colorado Boulder, 
University of Oxford, University of 
Portsmouth, University of Utah, 
University of Virginia, University 
of Washington, University of 
Wisconsin, Vanderbilt University, 
and Yale University.
\section*{Data Availability}

The data described in this paper are available to the public at
\url{https://data.sdss.org/sas/dr17/manga/spectro/specz/v3\_1\_1/1.0.1/}
\url{https://data.sdss.org/sas/dr17/manga/spectro/lensing/silo/v3\_1\_1/1.0.4/}
and are described at
\url{https://data.sdss.org/datamodel/files/MANGA\_SPECZ/DRPVER/SPECZVER/speczall.html}
\url{https://data.sdss.org/datamodel/files/MANGA\_SPECZ/DRPVER/SPECZVER/PLATE4/specz-RSS.html}
\url{https://data.sdss.org/datamodel/files/MANGA\_SPECZ/DRPVER/SPECZVER/PLATE4/specz-CUBE.html}
\url{https://data.sdss.org/datamodel/files/MANGA\_SPECTRO\_LENSING/silo/DRPVER/SILO_VER/silo_manga\_detections.html}
\url{https://data.sdss.org/datamodel/files/MANGA\_SPECTRO\_LENSING/silo/DRPVER/SILO_VER/PLATE/manga\_PLATE\_IFU\_stack\_data.html}.
\bibliographystyle{mnras}
\bibliography{tex/references}
\clearpage
\appendix
\label{sec:appendix}
\section{Tables and Figures}\label{appendix_tables}

\begin{sidewaystable}
\captionsetup{width=.8\linewidth}
\caption{Source-plane grading scheme used in geometry inspection. Column one lists the source-plane grade of assurance. Column two lists the requirements of the grade. Indications of lensed source features that are observed within the SN narrow-band can increase the grade of assurance assigned to a candidate. Any quality issues in either DAP maps or SDSS photometry can reduce the grade of assurance assigned to a candidate.}
\centering
\label{table:source_grading}
\setlength\tabcolsep{4pt}
\begin{tabular}{lp{15cm}}
 \hline\hline
Grade & Description \\
(1) & (2) \\
\hline
Grade A+ & The conditions of grade A are satisfied and lensed source features are observed in the SN narrowband image. \\
Grade A & Background emissions are assured, the predictions of the strong lensing regime are $\approx\ge1"$, the prediction of the strong lensing regime is robust across computations of the upper and approximated Einstein radius from both types of mass maps, and the candidate source is on or within the probable strong lensing regime. \\
Grade A- & All but one of the grade A conditions is satisfied, in which the failing property nearly satisfies the failed condition. \\
Grade B+ & The conditions of a grade A candidate are mostly satisfied. \\
Grade B & The conditions of a grade A candidate are partially satisfied. \\
Grade B- & The conditions of a grade A candidate are slightly less than partially satisfied. \\
Grade C+ & The candidate would have a grade of C if not for additional assurances. \\
Grade C & Either the quality of the detections is challenged, the strength of the overall strong lensing regime is challenged, the proximity of the candidate source to the strong lensing regime is challenged, or the candidate only satisfies a small fraction of the grade A conditions. \\
Grade C- & A feature of the candidate suggest lensing is not ruled out but no other assurances are demonstrated. \\
Grade X & The candidate background galaxy is ruled out as a candidate source. The most often justification is the nearest region of the candidate background galaxy is farther than several arcseconds away from the upper strong lensing regime. Contamination and reduction quality issues can also warrant the rejection of a candidate.

\\
\hline
\end{tabular}
\end{sidewaystable}

\begin{table}
\caption{Count of graded spectroscopic detections found within the co-added residuals, which more than one detection of candidate emission-lines may belong to a candidate background galaxy since the fibre bundle used in MaNGA observations may sample different regions of the background galaxy. Column one lists the spectra grade, which grades of A represent likely, B represent probable, and C represent possible background emissions. Column two lists the count of multi-line detections. Column three lists the count of single-line detections.}
\centering
\label{table:hits_in_sources}
\setlength\tabcolsep{4pt}
\begin{tabular}{lcc}
 \hline\hline
Grade & Multi-line Detections & Single-line Detections \\
(1) & (2) & (3)\\
\hline
A+ & 896 & 8 \\
A & 80 & 6 \\
A- & 49 & 3 \\
B+ & 15 & 1 \\
B & 7 & 2 \\
B- & 12 & 0 \\
C+ & 10 & 1 \\
C & 8 & 2 \\
C- & 1 & 4 \\
Total & 1,078 & 27 \\
\hline
\end{tabular}
\end{table}

\begin{table}
\caption{Count of candidate sources found within MaNGA target spectra. Column 1 lists the source grade, which grades of A represent likely, B represent probable, C represent possible, and X represent unlikely sources. Column two lists the count of candidate sources for each grade.}
\centering
\label{table:sourcecount}
\setlength\tabcolsep{4pt}
\begin{tabular}{lc}
 \hline\hline
Grade & Galaxy Count \\
(1) & (2)\\
\hline
A+ & 1 \\
A & 0 \\
A- & 7 \\
B+ & 0 \\
B & 8 \\
B- & 9 \\
C+ & 11 \\
C & 13 \\
C- & 45 \\
X & 363 \\
Total & 457 \\ \\
\hline
\end{tabular}
\end{table}

\begin{table*}
\centering
\caption{Table description of the primary extension of the MaNGA statistics summary file for each MaNGA target within the Specz VAC. Columns one, two, and three lists the name, data type, and description for each header, respectively.}
\label{table:speczall_primary}
\setlength\tabcolsep{2pt}
\def\arraystretch{0.3}
\begin{tabular}{lll}
\hline \hline
Name & Type & Description \\
(1) & (2) & (3)\\
\hline
AUTHOR & String &  Creater of specz files  \\
SOURCE & String &  Specz using MaNGA  \\
SPECZVER & String &  Version of Specz  \\
DRPVER & String &  Version of Data Reduction Pipeline  \\
RELEASE & String &  Release version  \\
SAMPTYPE & String &  Wavelength sampling type  \\
\\
\hline
\end{tabular}
\end{table*}

\begin{table*}
\centering
\caption{Table description of the \texttt{SUMMARY} extension of the MaNGA statistics summary file for each MaNGA target within the Specz VAC. Columns one, two, and three lists the name, data type, and description for each column, respectively.}
\label{table:speczall_summary}
\setlength\tabcolsep{2pt}
\def\arraystretch{0.3}
\begin{tabular}{lll}
\hline \hline
Name & Type & Description \\
(1) & (2) & (3) \\
\hline
MANGA\_ID & String &  MaNGA ID \\
FILETYPE & String &  File is either RSS(fibers) or CUBE(Spaxels) \\
PLATE & Integer &  Target plate \\
IFU & Integer &  Target IFU \\
NSA\_Z & Float &  NSA redshift \\
NSPEC & Integer &  Number of spectra within MaNGA target \\
OBJDEC\textsuperscript{a} & Float &  Object declination \\
OBJRA\textsuperscript{a} & Float &  Object right ascension \\
IFUDEC\textsuperscript{a} & Float &  IFU declination \\
IFURA\textsuperscript{a} & Float &  IFU right ascension \\
COMP & Float &  Percent of redshifts recorded \\
MEAN\_Z & Float &  Mean redshift \\
STD\_Z & Float &  Standard deviation of redshifts \\
SKEW & Float &  Skewness \\
MEDIAN\_Z & Float &  Median redshift \\
IQR & Float &  Interquartile range of redshifts \\
Q25 & Float &  25th quartile \\
Q75 & Float &  75th quartile \\
R0 & Float &  Radius of inner sample \\
COMP0 & Float &  Percent of redshifts within R0 \\
MEAN\_Z0 & Float &  Mean redshift within R0 \\
STD\_Z0 & Float &  Standard deviation within R0 \\
SKEW0 & Float &  Skewness within R0 \\
MEDIAN\_Z0 & Float &  Median redshift within R0 \\
IQR0 & Float &  Interquartile range of redshifts within R0 \\
Q250 & Float &  25th quartile within R0 \\
Q750 & Float &  75th quartile within R0 \\
\\
\hline
\end{tabular}
\end{table*}

\begin{table*}
\centering
\caption{\label{table:specz_vac_table_primary}Table description of the primary extension of the MaNGA Specz VAC fits file. Columns one, two, and three lists the name, data type, and description for each header, respectively.}
\setlength\tabcolsep{2pt}
\def\arraystretch{0.3}
\begin{tabular}{lll}
\hline \hline
Name & Type & Description \\
(1) & (2) & (3)\\
\hline
RELEASE & String & SDSS data release  \\
DRPVER & String & DRP version  \\
SPECZVER & String & Specz code version  \\
FILETYPE & String & Data structure used to compute Specz  \\
MANGA\_ID & String & Target id  \\
PLATE & Integer & Target plate  \\
IFUDSGN & Integer & Target IFU  \\
NSA\_Z & Float & Target NSA redshift  \\
OBJRA\textsuperscript{a} & Float & Target right ascension in degrees  \\
OBJDEC\textsuperscript{a} & Float & Target declination in degrees  \\
R0\textsuperscript{b} & Float & High signal-to-noise inner sample radius  \\
BIN\_SIZE & Integer & Sample size used to compute mean z  \\
MEANZ\_R0 & Float & Mean z within R0  \\
SIGZ\_R0 & Float & Sigma of z within R0  \\
Z\_TOL & Float & Maximum delta redshift allowed  \\
\\
\hline
\end{tabular}
\end{table*}

\begin{table*}
\centering
\caption{\label{table:specz_vac_table_redshifts}Table description of the \texttt{REDSHIFT} extension of the MaNGA Specz VAC fits file. Columns one, two, and three lists the name, data type, and description for each column, respectively.}
\setlength\tabcolsep{2pt}
\def\arraystretch{0.3}
\begin{tabular}{lll}
\hline \hline
Name & Type & Description \\
(1) & (2) & (3) \\
\hline
INDEX\_ROW & Integer (Array) & Row index in MaNGA CUBE or RSS file  \\
INDEX\_COL & Integer (Array) & Column index in MaNGA CUBE file (for spaxels only) \\
SPECZ & Float (Array) & Spectroscopic redshifts  \\
SPECZ\_ZNUM & Float (Array) & Extension of specz  \\
SPECZ\_ERR & Float (Array) & Spectroscopic redshifts error  \\
SPECZ\_RCHI2 & Float (Array) & Spectroscopic redshifts reduced chi2  \\
MEAN\_X\textsuperscript{a} & Float (Array) & Mean of xpos from galaxy center  \\
MEAN\_Y\textsuperscript{a} & Float (Array) & Mean of ypos from galaxy center  \\
MEAN\_R\textsuperscript{a} & Float (Array) & Mean radius from galaxy center \\
MEAN\_SN & Float (Array) & Mean spectra signal-to-noise  \\
\\
\hline
\end{tabular}
\end{table*}

\begin{table*}
\centering
\caption{Table description of the primary extension of the summary file contained within the MaNGA lens VAC. Columns one, two, and three lists the name, data type, and description for each header, respectively.}
\label{table:vac_table_overview}
\setlength\tabcolsep{2pt}
\def\arraystretch{0.3}
\begin{tabular}{lll}
\hline \hline
Name & Type & Description \\
(1) & (2) & (3) \\
\hline
PROJECT & String & Lens search project  \\
AUTHORS & String & SILO project creators  \\
INSP & String & Detection inspector  \\
SCANNED & String & SDSS surveys scanned by SILO  \\
DRPVER & String & Version of MaNGA data reduction pipeline  \\
SILO\_VER & String & Version of SILO project used  \\
RELEASE & String & SDSS data release version  \\
SAMPLING & String & Wavelength sampling of spectra  \\
SPECTYPE & String & Type of spectra searched  \\
OIIB\textsuperscript{a} & Float & Restframe wavelength of OIIb  \\
OIIA\textsuperscript{a} & Float & Restframe wavelength of OIIa  \\
HID\textsuperscript{a} & Float & Restframe wavelength of HId  \\
HIC\textsuperscript{a} & Float & Restframe wavelength of HIc  \\
HIB\textsuperscript{a} & Float & Restframe wavelength of HIb  \\
OIIIB\textsuperscript{a} & Float & Restframe wavelength of OIIIb  \\
OIIIA\textsuperscript{a} & Float & Restframe wavelength of OIIIa  \\
NIIB\textsuperscript{a} & Float & Restframe wavelength of NIIb  \\
HIA\textsuperscript{a} & Float & Restframe wavelength of HIa  \\
NIIA\textsuperscript{a} & Float & Restframe wavelength of NIIa  \\
SIIB\textsuperscript{a} & Float & Restframe wavelength of SIIb  \\
SIIA\textsuperscript{a} & Float & Restframe wavelength of SIIa  \\
\\
\hline
\end{tabular}
\end{table*}

\begin{table*}
\centering
\caption{Table description of the \texttt{CANDIDATE\_OVERVIEW} extension of the summary file contained within the MaNGA lens VAC. Columns one, two, and three lists the name, data type, and description for each column, respectively.}
\label{table:vac_table_source}
\setlength\tabcolsep{2pt}
\def\arraystretch{0.3}
\begin{tabular}{lll}
\hline \hline
Name & Type & Description \\
(1) & (2) & (3) \\
\hline
BACKGROUND\_ID & Integer & Assinged background id  \\
SDSS\_TARGET\_NAME & String & SDSS RA+DEC name of target  \\
MANGA\_ID & String & MaNGA target identifier  \\
PLATE & Integer & MaNGA plate  \\
IFU & Integer & MaNGA integral field unit  \\
NSA\_ZF & Float & Target redshift from NSA  \\
MEAN\_ZF\_INNER & Float & Mean inner z from Specz VAC  \\
SIGMA\_ZF\_INNER & Float & Sigma inner z from Specz VAC  \\
RA\_IFU\textsuperscript{a} & Float & Right ascension of MaNGA IFU  \\
DEC\_IFU\textsuperscript{a} & Float & Declination of MaNGA IFU  \\
MEAN\_ZB & Float & Mean z of detections  \\
SIGMA\_ZB & Float & Sigma z of detections  \\
SOURCE GRADE & String & Inspection grade of candidate lens  \\
SOURCE COMMENT & String & Inspection comment of candidate lens  \\
NEAREST\_RATIO & Float & Ratio of nearest source edge to upper limit ER  \\
ER\_JAM\textsuperscript{b} & Float (Array) & Lower, best, upper Einstein radius using DAP  \\
ER\_FIREFLY\textsuperscript{b} & Float (Array) & Lower, best, upper Einstein radius using DAP  \\
DETECTION\_COUNT & Integer & Count of graded detections  \\
\\
\hline
\end{tabular}
\end{table*}

\begin{table*}
\centering
\caption{Table description of the \texttt{DETECTION} extension of the summary file contained within the MaNGA lens VAC. Columns one, two, and three lists the name, data type, and description for each column, respectively.}
\label{table:vac_table_detection}
\setlength\tabcolsep{2pt}
\def\arraystretch{0.3}
\begin{tabular}{lll}
\hline \hline
Name & Type & Description \\
(1) & (2) & (3) \\
\hline
MANGA\_ID& String & MaNGA target id  \\
DETECTION\_ID& Integer & Detection identifier for MaNGA target  \\
EMLINE\_SCAN\_TYPE& String & Single-line=OII(b, a), Multi-line=2+ lines  \\
FOREGROUND\_Z & Float & Mean redshift across stacked RSS  \\
DETECTION\_Z & Float & Redshift of background candidate  \\
N\_EMLINES\_SN\_GE4 & Integer & Number of emission-lines detected with SN>=4  \\
QUADATURE\_SUM\_SN\_GE3 & Float & Quadrature sum of emission-lines with SN>=3  \\
MEAN\_X\textsuperscript{a} & Float & Mean x of stacked fibers  \\
MEAN\_Y\textsuperscript{a} & Float & Mean y of stacked fibers  \\
MEAN\_R\textsuperscript{a} & Float & Mean stacked radius from galaxy center  \\
FIBER\_POSITION& String & Fiber number in IFU and dither position  \\
STACKED\_COUNT& Float & Count of stacked RSS residuals  \\
STACKED\_RSS& Float (Array) & Row indexs of stacked RSS residuals  \\
SPECTRA\_GRADE& String & Grade of spectra assurances of source candidate  \\
COMMENT& String & Comment of assuring/non-assuring features  \\
\\
\hline
\end{tabular}
\end{table*}

\begin{table*}
\centering
\caption{Table description of the \texttt{EMISSION\_LINE\_ANALYSIS} extension of the summary file contained within the MaNGA lens VAC. Columns one, two, and three lists the name, data type, and description for each column, respectively.}
\label{table:vac_table_analysis}
\setlength\tabcolsep{2pt}
\def\arraystretch{0.3}
\begin{tabular}{lll}
\hline \hline
Name & Type & Description \\
(1) & (2) & (3)\\
\hline
DETECTION\_ID & Integer & Detection identifier for MaNGA target  \\
NAME & String & Name of emission-line  \\
INDEX\_IN\_SPECTRA & Integer & Index emission-line is located in spectra  \\
EM\_WAVE\textsuperscript{a} & Float & Observed-frame wavelength of emission-line  \\
SN & Float & Convolved signal-to-noise of emission-line  \\
GAUSS\_FIT\_REPORTED & BOOLEAN & Gauss fit reported if SN>=3  \\
GAUSS\_WAVE\textsuperscript{a} & Float (Array) & Wavelength(s) center of model w. -/+97.5 Q.  \\
GAUSS\_BASE\_HEIGHT\textsuperscript{a} & Float (Array) & Gaussian model base height w. -/+97.5 Q.  \\
GAUSS\_AMPLITUDE\textsuperscript{a} & Float (Array) & Gaussian model amplitude w. -/+97.5 Q.  \\
GAUSS\_SIGMA\textsuperscript{a} & Float (Array) & Gaussian model sigma w. -/+97.5 Q.  \\
RCHI2\_SAMPLE & Float & Reduced chi\^2 of Gauss fit to sample  \\
NDOF\_SAMPLE & Integer (Array) & Degree of freedom of RCHI2\_SAMPLE  \\
RCHI2\_3SIG & Float & Reduced chi\^2 of Gauss fit within 3 sigma  \\
NDOF\_3SIG & Integer (Array) & Degree of freedom of RCHI2\_3SIG  \\
SAMPLE\_SIZE & Integer (Array) & Sample size  \\
MODEL\_WAVE\_BASE\textsuperscript{a} & Float (Array) & Wavelength base of Gaussian model  \\
GAUSS\_MODEL\textsuperscript{b} & Float (Array) & Gaussian model of residual flux  \\
FITTED\_RESIDUAL\_FLUX\textsuperscript{b} & Float (Array) & Residual flux segment used in Gaussian fit  \\
FITTED\_IVAR\_RESCALED & Float (Array) & Rescaled inverse varience used in Gauss fit  \\
AND\_MASK & Integer (Array) & Co-added AND\_MASK of spectra sample  \\
OR\_MASK & Integer (Array) & Co-added OR\_MASK of spectra sample  \\
\\
\hline
\end{tabular}
\end{table*}

\begin{table*}
\centering
\caption{Table description of the primary extension for the MaNGA lens VAC file. Columns one, two, and three lists the name, data type, and description for each header, respectively.}
\label{table:vac_table_primary_stack}
\setlength\tabcolsep{2pt}
\def\arraystretch{0.3}
\begin{tabular}{lll}
\hline \hline
Name & Type & Description \\
(1) & (2) & (3)\\
\hline
RELEASE & String & SDSS data release  \\
DRPVER & String & DRP version  \\
SPECZVER & String & Specz code version  \\
FILETYPE & String & Data structure used to compute Specz  \\
MANGA\_ID & String & Target id  \\
PLATE & Integer & Target plate  \\
IFUDSGN & Integer & Target IFU  \\
NSA\_Z & Float & Target NSA redshift  \\
OBJRA\textsuperscript{a} & Float & Target right ascension in degrees  \\
OBJDEC\textsuperscript{a} & Float & Target declination in degrees  \\
R0\textsuperscript{b} & Float & High signal-to-noise inner sample radius  \\
BIN\_SIZE & Integer & Sample size used to compute mean z  \\
MEANZ\_R0 & Float & Mean z within R0  \\
SIGZ\_R0 & Float & Sigma of z within R0  \\
Z\_TOL & Float & Maximum delta redshift allowed  \\
\\
\hline
\end{tabular}
\end{table*}
\begin{table*}
\centering
\caption{Table description of the \texttt{INDIVIDUAL\_EXPOSURE\_REDUCTION} extension of the MaNGA lens VAC file. Columns one, two, and three lists the name, data type, and description for each column, respectively.}
\label{table:vac_table_rss}
\setlength\tabcolsep{2pt}
\def\arraystretch{0.3}
\begin{tabular}{lll}
\hline \hline
Name & Type & Description \\
(1) & (2) & (3) \\
\hline
INDEX\_ROW & Integer (Array) & Row index in MaNGA CUBE or RSS file  \\
INDEX\_COL & Integer (Array) & Column index in MaNGA CUBE file (for spaxels only) \\
SPECZ & Float (Array) & Spectroscopic redshifts  \\
SPECZ\_ZNUM & Float (Array) & Extension of specz  \\
SPECZ\_ERR & Float (Array) & Spectroscopic redshifts error  \\
SPECZ\_RCHI2 & Float (Array) & Spectroscopic redshifts reduced chi2  \\
MEAN\_X\textsuperscript{a} & Float (Array) & Mean of xpos from galaxy center  \\
MEAN\_Y\textsuperscript{a} & Float (Array) & Mean of ypos from galaxy center  \\
MEAN\_R\textsuperscript{a} & Float (Array) & Mean radius from galaxy center \\
MEAN\_SN & Float (Array) & Mean spectra signal-to-noise  \\
\\
\hline
\end{tabular}
\end{table*}

\begin{table*}
\centering
\caption{Table description of the \texttt{STACKED\_REDUCTION} extension of the MaNGA lens VAC file. Columns one, two, and three lists the name, data type, and description for each column, respectively.}
\label{table:vac_table_stack}
\setlength\tabcolsep{2pt}
\def\arraystretch{0.3}
\begin{tabular}{lll}
\hline \hline
Name & Type & Description \\
(1) & (2) & (3) \\
\hline
FIBER\_POSITION & String & Fiber number and dither position  \\
RSS\_INDX & Integer (Array) & Row-stacked index(s) of spectra in stack  \\
EXPNUM & Integer (Array) & Stacked exposure numbers  \\
SET & Integer (Array) & Stacked dither pattern iterations  \\
SEEING\textsuperscript{a} & Float (Array) & Seeing of stacked RSS  \\
MEAN\_X\textsuperscript{a} & Float & Stacked RSS mean x position  \\
MEAN\_Y\textsuperscript{a} & Float & Stacked RSS mean y position  \\
MEAN\_R\textsuperscript{a} & Float & Stacked RSS mean r position  \\
MEAN\_THETA & Float & Stacked RSS mean angle from East  \\
MEAN\_SPECZ & Float & Mean redshift of stacked RSS  \\
MEAN\_XPOS\textsuperscript{a} & Float (Array) & Stacked mean wavelength-dependent x position  \\
MEAN\_YPOS\textsuperscript{a} & Float (Array) & Stacked mean wavelength-dependent y position  \\
MEAN\_RPOS\textsuperscript{a} & Float (Array) & Stacked mean wavelength-dependent radius  \\
WAVE Angstroms & Float (Array) & Wavelength of spectra  \\
AND\_MASK & Integer (Array) & Andmask of stacked spectra  \\
OR\_MASK & Integer (Array) & Ormask of stacked spectra  \\
SRESFLUX\textsuperscript{b} & Float (Array) & Stacked residual spectra of flux-model  \\
SIVAR & Float (Array) & Stacked rescaled inverse variance  \\
SSN\_SPEC & Float (Array) & Gaussian convolved SN of stack  \\
SO2SN\_SPEC & Float (Array) & OII Gaussian convolved SN of stack  \\
\\
\hline
\end{tabular}
\end{table*}

\begin{table*}
\centering
\caption{This table describes each ImageHDU extension contained within the MaNGA lens VAC file. Columns one and two lists the name and description for each column, respectively.}
\label{table:vac_table_spaxels}
\setlength\tabcolsep{2pt}
\def\arraystretch{0.3}
\begin{tabular}{ll}
\hline \hline
Name & Description \\
(1) & (2)\\
\hline
WAVE\textsuperscript{a} & Wavelength vector  \\
SPECRES & Median spectral resolution vs wavelength  \\
SPECRESD & Standard deviation (1-sigma) of spectral resolution vs wavelength  \\
PRESPECRES & Median pre-pixel spectral resolution vs wavelength \\
PRESPECRESD & Standard deviation of pre-pixel spectral resolution vs wavelength  \\
CUBE\_INDIVIDUAL\_FLUX\textsuperscript{b} & Flux of individual spectra  \\
CUBE\_INDIVIDUAL\_MODEL\textsuperscript{b} & Model of individual spectra  \\
CUBE\_INDIVIDUAL\_RESIDUALS\textsuperscript{b} & Residuals of flux-foreground model \\
CUBE\_INDIVIDUAL\_IVAR\_RESCALED & Re-scaled inverse variance of flux-foreground residuals  \\
CUBE\_INDIVIDUAL\_GAUSS\_SN & Gaussian convolved SN of residuals  \\
CUBE\_INDIVIDUAL\_GAUSS\_O2SN & OII Gaussian convolved SN of residuals  \\
CUBE\_STACKED\_RESIDUALS\textsuperscript{b} & Co-added residuals  \\
CUBE\_IVAR\_RESCALED\_STACKED & Inverse variance of co-added residuals  \\
CUBE\_STACKED\_GAUSS\_SN & Gaussian convolved SN of co-added residuals  \\
CUBE\_STACKED\_GAUSS\_O2SN & OII Gaussian convolved SN of co-added residuals  \\
\\
\hline
\end{tabular}
\end{table*}

\clearpage

\section{Computation of the Upper Limit of the Total Mass-Density Map From Dynamics}\label{jamdynamics}

One approach to approximate the total mass map is to fit a mass profile, inclination, anisotropy, and position angle to the stellar dynamics. SILO uses \texttt{JamPy} to fit a stellar velocity map to a model of the galaxy potential. In particular, \texttt{jam\_axi\_proj} solves the Jeans~\citep{1922MNRAS..82..122J} equations to generate a model of the second velocity moment:
\begin{equation}
RMS_{LOS} = \sqrt{v_{LOS}^2 + \sigma_{LOS}^2}
\end{equation}
of either a spherically or cylindrically axis JAM based upon a trial radial anisotropy, black hole (BH) mass at the centre of the galaxy, inclination, galaxy density profile, and the $RMS_{LOS}$ map. The trial input parameters can then be varied until the JAM model best matches observations. 

The galaxy density profile used by \texttt{JamPy} is expressed in the form of Multi Gaussian Expansion (MGE) parameterizations~\citep{1994A&A...285..723E} of either a radial density or surface density profile, in which the \texttt{MgeFit}~\citep{2002MNRAS.333..400C} package can be used to fit an MGE to a density profile. There are many advantages to using an MGE fit of the density profile. Section~2.2 of~\cite{2002MNRAS.333..400C} demonstrates how the observed axial ratios ($q_j$) of a surface density MGE can be used to deproject the surface density MGE into a radial density MGE. Section~2.1 of~\cite{2002MNRAS.333..400C} demonstrates how a PSF of the seeing can be incorporated as an MGE in the luminosity surface density MGE to convolve the projection of the light profile during the fit to imaging. Finally, equation 11 from~\cite{2013MNRAS.432.1709C} demonstrates how the light enclosed within a radius (R) along the LOS can be computed from a luminosity surface density MGE given the total luminosity of each Gaussian. The values of equation 11 from~\cite{2013MNRAS.432.1709C} can also be switched for the total mass of each Gaussian to compute the LOS mass enclosed within R given the mass-surface density MGE. Thus, MGEs are not only useful to project the mass enclosed within a sphere that impacts stellar dynamics but can also be used to project the mass enclosed within R along the LOS.

\subsection{Computation of the Stellar Mass-Density Profile}\label{apdx_stellar_mass_profile}

The distribution of the stellar mass-density profile can be projected by the light profile of the galaxy multiplied by a stellar-mass to a stellar light ratio ($M_*/L_*$). However, SILO must first retrieve the SDSS photometry and prepare the photometry for an MGE fit of the light profile. The coordinates, seeing PSF in the SDSS r-band (i.e. red filter), and the radius that contains $90\%$ of the Petrosian fitted flux in the r-band ($R_{90}$) is obtained for both the target and nearby celestial objects from the \texttt{PhotoObj} database and retrieved by the SDSS Sky Server\footnote{SDSS SKY SERVER: \url{http://skyserver.sdss.org}}. SILO then obtains a cutout of the target within twice the semi-major axis of $R_{90}$ from SDSS Legacy photometry~\citep{2001ASPC..238..269L} in the  r-band, which image file is obtained from SDSS \texttt{frames} files located within the DR17 PhoObj directory\footnote{PhotoObj: \url{https://data.sdss.org/sas/dr17/eboss/photoObj}} on the SAS. However, nearby objects can cause the measurement of $R_{90}$ to underrepresent the light profile. Thus SILO will instead retrieve a photometry cutout within twice the IFU radius if $R_{90}$ is below five arcseconds. 

However, other galaxies and stars can contaminate the image and thus impact target modeling. Thus SILO applies a methodology to mask where the observed surface brightness from contaminating sources is brighter than the target's observed surface brightness in the same region as such contamination can significantly impact target modeling. SILO also adds a constraint that the contamination must be brighter than the background sky to prevent masking beyond the significant region of the source. Inspection of contamination fainter than either condition reveals negligible impact on SILO's fits of the target flux.

In particular, SILO first retrieves the r-band Petrosian, De Vaucouleurs, Exponential, and PSF fitted flux data from PhotoObj for all objects within the field of view of the image cutout. Non-target objects that have a Petrosian-derived $SN<3$ are excluded since the observed surface brightness for these objects is often insufficient for proper SDSS modeling, which is used in the masking process explained below. The exclusion is justified since these faint objects insignificantly impact SILO's fit of the target flux. Galaxies that reside within the target radius containing $50\%$ of the Petrosian fitted flux are also excluded since these objects are often merging with the target or represent a second measurement of the target. In addition, the observed surface brightness of foreign galaxies is often less than the observed surface brightness of the target within this region. The PSF flux of nearby stars is then projected as a single point onto an observed surface brightness map. In addition, the observed surface intensity from the weighted combination of the best fitted De Vaucouleurs and Exponential models from SDSS are also projected onto the observed surface brightness map for all non-target galaxies. A second observed surface brightness map is also created for the target using the weighted combination of the best fitted De Vaucouleurs and Exponential models from SDSS. Both observed surface brightness maps are then convolved with a two-dimensional PSF map constructed from the seeing data contained within SDSS's \texttt{psField} files using the method\footnote{PSF method: \url{https://www.sdss.org/dr17/imaging/images/\#psf}} developed by SDSS. The background threshold is then measured as 1.5 times the interquartile range over the seventy-fifth percentile of the sky within a ten-arcsecond region on the image that contains the lowest median background. The image mask is then applied where the cumulative observed surface brightness of non-target objects exceeds the observed surface brightness of the target when comparing observed surface brightness maps, which masking is limited to where the observed surface brightness of the contamination also exceeds the background threshold.

SILO uses the \texttt{find\_galaxy} function included in the \texttt{MgeFit} package to measure the position angle (PA), major axis, ellipticity, and center of the target in the image. These values from \texttt{find\_galaxy}, the photometry, and the mask are used as input into the \texttt{sectors\_photometry} function from the \texttt{MgeFit} package, in which \texttt{sectors\_photometry} extracts spatially binned photometric measurements from the galaxy image. The photometric measurements, ellipticity value from \texttt{find\_galaxy}, the pixel scale in arcseconds, and the PSF of the red image in units of pixels is next used in the \texttt{mge\_fit\_sectors\_regularized} function to fit up to 20 Gaussians to the photometric measurements. The MGE returned by \texttt{mge\_fit\_sectors\_regularized} has been de-convolved to mitigate the impact of seeing on smoothing the steep gradient of the light profile within the center of the target. 

The \texttt{jam\_axi\_proj} function uses an MGE with each Gaussian $\sigma_j$ in arcseconds and each Gaussian magnitude in peak solar luminosity per square parsec ($L_{\odot}\mathrm{pc^{-2}}$) to constrain the stellar distribution. In preparation to rescale the fitted MGE, the solar luminosity of the target in the r-band is obtained from:
\begin{equation}
L_{\odot} = 10^{0.4(AB_{\odot} - AB_{\mathrm{galaxy}})}
\end{equation}
where $AB_{\mathrm{galaxy}}$ is the S\'ersic absolute magnitude in the r-band recorded as \texttt{NSA\_SERSIC\_ABSMAG} in the DRPALL file\footnote{DRPALL file: \url{https://dr17.sdss.org/sas/dr17/manga/spectro/redux/v3_1_1/drpall-v3_1_1.fits}} of SDSS and $AB_{\odot}$ is the AB magnitude of the sun in the SDSS r-band~\citep{2003ApJ...592..819B}. The total flux of the fitted MGE in units of nanomaggies is then re-scaled by the ratio of $L_{\odot}$ to the sum of the total flux from all Gaussians to yield the total luminosity ($L_j$) of each Gaussian. Each $L_j$ is then converted into the central surface brightness ($L_{\odot}\mathrm{pc^{-2}}$) using equation ten of~\cite{2013MNRAS.432.1709C} and $\sigma_j$ is re-scaled to arcseconds.

The \texttt{jam\_axi\_proj} function also uses a total mass density MGE to fit the $RMS_{LOS}$ map. The stellar mass density is approximated by multiplying the luminosity MGE by a trial $M_*/L_*$. The stellar mass MGE is combined with a DM halo MGE to approximate the total density MGE, which computation of the DM halo MGE is described in the following section.

\subsection{Computation of the Density Profile of the Dark Matter Halo}\label{dmh}

An MGE fit of the physical DM halo is obtained by using an NFW equation to project the physical density across radius once the total physical DM halo mass and the DM halo concentration are quantified. The total physical mass of the DM halo is projected by multiplying the trial total stellar mass with the DM halo mass to stellar M/L predicted using equation 6 and Table 2 from~\cite{2020A&A...634A.135G}, which the latter is predicted based upon the $\Lambda CDM$ DUSTGRAIN-pathfinder simulation~\citep{2018MNRAS.481.2813G}. The DM halo mass is then used as input into the \texttt{conc\_NFWmodel} function from \texttt{halotools} package~\citep{2017AJ....154..190H} to compute the NFW halo concentration by using a power-law fit to the concentration-mass relation from equations 12 and 13 of~\citet{2014MNRAS.441.3359D}. SILO then uses the \texttt{mass\_density} function from the \texttt{halotools} package to project the radial density of a symmetric NFW halo in a flat $\Lambda CDM$ cosmology using the total halo physical mass and halo concentration as input, in which SILO re-scales the density from units of $\rho \mathrm{ h^{-3}Mpc^{-3}}$ to units of $\rho \mathrm{pc^{-3}}$. The density profile is then fitted by the \texttt{mge\_fit\_1d} function from the \texttt{MgeFit} package to obtain an MGE with a Gaussian central surface density in units of $M_{\odot}\mathrm{pc^{-2}}$. Thus only the Gaussian dispersion of the MGE needed to be converted to arcseconds as input into \texttt{jam\_axi\_proj}.

\subsection{Preparation of the Stellar Velocity Maps}

The \texttt{jam\_axi\_proj} function can use a $RMS_{LOS}$ map in the fit of the total density profile, which the $RMS_{LOS}$ map can be computed from adding in quadrature a $v_{LOS}$ map and a $\sigma_{LOS}$ map. However, SILO does not use the $\sigma_{LOS}$ spaxel map from the \texttt{SPX-MILESHC-MASTARSSP} file since measurements of $\sigma_{LOS}$ can be significantly biased in low SN spaxels (see Section~\ref{cjam}). Instead, the $\sigma_{LOS}$ map selected by SILO is obtained from the \texttt{VOR10-MILESHC-MASTARSSP} file since the Voronoi binned measurements of $\sigma_{LOS}$ are evaluated with a $SN\ge10$. The \texttt{VOR10-MILESHC-MASTARSSP} still contains the binned measurements across spaxels but also contains a $\sigma_{LOS}$ correction map to subtract systematic biases from the $\sigma_{LOS}$ map. The $v_{LOS}$ maps are obtained both from the SPX-MILESHC-MASTARSSP and VOR10-MILESHC-MASTARSSP files, which the former is only used in computing the position angle (PA) of the $v_{LOS}$ map since the smoother spaxel resolution yields better orientation fits than the Voronoi bins.

The maps from the DAP can contain quality issues caused by intervening objects and other forms of contamination. Thus the velocity maps must be prepared to improve the fit of the mass-density profile to the derived $RMS_{LOS}$ map, which preparation is performed across the spaxels of the maps prior to extracting $RMS_{LOS}$ computed from each Voronoi bin in order to preserve the area representation of the sampling. In particular, a spaxel is masked if the inverse variance of the spaxel is zero for either the $\sigma_{LOS}$ or $v_{LOS}$ values obtained from the DAP maps, if the coordinates of the spaxel are located within a masked region of the photometry, or if the values are non-finite. During an inspection of the VOR10-MILESHC-MASTARSSP maps, it was also noticed that bright distant stars can still contaminate the measured $\sigma_{LOS}$ on the side of the target that is directed towards the star. In addition, inspection revealed that spectra contributions from faint regions of additional foreground galaxies can significantly bias kinematic measurements.  Unfortunately, neither DRP nor DAP currently has a method that robustly masks these contamination regions. To mitigate this issue, SILO applies a masking method that is described along with the related processes throughout this section.

The \texttt{jam\_axi\_proj} function expects the coordinates of the $RMS_{LOS}$ to be aligned along the major axis of the galaxy and velocity systematics to be corrected. In preparation to compute the PA used to rotate the $RMS_{LOS}$ coordinates, SILO subtracts $v_{LOS}$ by its median to mitigate systematic biases in the LOS velocity. The PA is then computed from \texttt{fit\_kinematic\_pa}, which uses the method listed in Appendix C of~\citet{2006MNRAS.366..787K} to determine the PA of the median subtracted $v_{LOS}$ map. The \texttt{fit\_kinematic\_pa} function further computes the systematic bias in the median subtracted $v_{LOS}$ map, which is used to correct for any systematic bias in the velocity. Masks are next applied to any velocity above 400 km\,s$^{-1}$ in the systematic corrected $v_{LOS}$ map or the $\sigma_{LOS}$ map, which velocity is un-physical for the kinematics of galaxies.

Since $\sigma_{LOS}$ measurements are easily biased by low SN measurements or contaminants, SILO utilizes the prior mask but applies more methods to properly reject poor $\sigma_{LOS}$ measurements. In particular, $\sigma_{LOS}$ measurements above 400 km\,s$^{-1}$ are rejected as un-physical. Any $\sigma_{LOS}$ measurement with a mean g-band-weighted pixel $SN < 5$ in the Voronoi binned spectra is masked since test from~\cite{2019AJ....158..231W} reveals lower SN measurements are unreliable, and our inspection of the $\sigma_{LOS}$ maps reveals these measurements often reside in regions where the brightness of the sky background is equivalent to the target brightness. Next, $\sigma_{LOS}$ measurements that are 10 km\,s$^{-1}$ beyond both the interquartile range and the continuous distribution of $\sigma_{LOS}$ measurements are masked since such measurements are obviously biased beyond typical measurements. Voronoi bins that contain more than 10\% of masked spaxels are then completely masked, which percentage threshold is chosen to avoid masking bins where sometimes several reasonable measurements are masked by photometry or reduction masking processes.

Inspection of the $\sigma_{LOS}$ maps for a hundred targets revealed tighter masks are required to remove bins with obvious bias that reside within the continuous distribution of the $\sigma_{LOS}$ measurements. To do this, SILO first computes an elliptical radius ($R_e$) , defined by a major to minor axis ratio and position angle that approximates contours of constant $\sigma_{LOS}$ within the $\sigma_{LOS}$ map of the target. The $R_e$ is approximated using the PA measurement and the major to minor axis ratio of the SDSS Sersic fit of the target. SILO then iterates over $R_e$ to determine the median $\sigma_{LOS}$ located between $R_e$ and $R_e+1"$. SILO then masks $\sigma_{LOS}$ measurements that are below or above 1.5 times the interquartile range (IQR) from the twenty-fifth percentile (q25) or seventy-fifth percentile (q75) of the $\sigma_{LOS}$ minus median sample, respectively. Since the error in $\sigma_{LOS}$ strongly correlates with the SN of each measurement, evaluating IQR, q75, and q25 across the median-subtracted $\sigma_{LOS}$ yields a global masking threshold that is more likely to reject poor measurements typically located farther from the target center. Finally, $\sigma_{LOS}$ measurements below 50 km\,s$^{-1}$ are rejected since DAP test reveal such measurements are typically unreliable within MaNGA kinematics~\citep{2019AJ....158..231W}. Voronoi bins that contain more than 10\% of masked spaxels are then completely masked, which percentage threshold is chosen to avoid masking bins where sometimes several reasonable measurements are masked because of variance introduced by the median map.

Unfortunately, the additional $\sigma_{LOS}$ masking methods would mask most of the sample from rotationally-dominated targets where most $\sigma_{LOS}$ measurements above 50 km\,s$^{-1}$ may only reside within a few arcseconds from the target center. As a solution, SILO evaluates if the $v_{LOS}$ can be used instead of a $RMS_{LOS}$ for JAM modeling of rotationally-dominated systems. In particular, SILO will instead provide $v_{LOS}$ and its mask to JAM (including instructing JAM to compute the radial velocities) if any of the following conditions are met:
\begin{enumerate}
\item The $\sigma_{LOS}$ mask completely rejects the inner arcsecond sample from the target center, which infers $\sigma_{LOS}$ is typically below the 50 km\,s$^{-1}$ mask threshold.
\item The ninety percentile (q90) of un-masked $v_{LOS}$ is $\sqrt{10}$ greater than q90 of the un-masked $\sigma_{LOS}$ sample, inferring that $\sigma_{LOS}$ only contributes about a tenth of the velocities added in quadrature to obtain the $RMS_{LOS}$ map and thus negligibly contributes about a twentieth to $RMS_{LOS}$. The purpose of comparing q90 measurements instead of the maximum in each map is to better represent typically larger measurements.
\item The sample size of un-masked $v_{LOS}$ is three times greater than the sample size of un-masked $\sigma_{LOS}$ within the half-light radius evaluated by a Sersic profile fit from SDSS. This step helps to reduce the risk that poorer $\sigma_{LOS}$ measurements can yield a masked-dominated sample.
\end{enumerate}

Simulations and comparisons between multiple observations of the same target also infer that kinematic uncertainties should be increased by a factor of $log10(SN)$ where the binned $SN > 10$~\citep{2019AJ....158..231W}. SILO thus includes this re-scaling into the kinematic uncertainties prior to computing $RMS_{LOS}$ from the systematic corrected $v_{LOS}$ map and the $\sigma_{LOS}$ map. The un-masked Voronoi $RMS_{LOS}$ measurements are then extracted from an arbitrary spaxel belonging to each bin.

\subsection{Computation of the Total Density Profile}\label{densitycomputation}

SILO uses the AdaMet~\citep{2013MNRAS.432.1709C} optimizer to fit a total density profile to the $RMS_{LOS}$ map, which optimizer uses an Adaptive Metropolis algorithm~\citep{bj/1080222083} to perform the fit. SILO provides a trial function to the optimizer that:
\begin{itemize}
\item Creates a DM halo MGE as specified in Section~\ref{dmh}.
\item Project the stellar mass MGE by multiplying the stellar light MGE by a trail $M_*/L_*$.
\item Rotates the PA corrected coordinates of the $RMS_{LOS}$ map by the trail offset angle.
\item Combines the stellar density MGE and DM density MGE to project a total density MGE.
\item Converts a trial internal axis rotation parameter ($q$) into a projection of the galaxy inclination ($i$) by:
\begin{equation}
i = atan\left(\sqrt{\frac{1 - min(q_j)^2}{min(q_j)^2 - q^2}}\right)\label{eq:inclination}
\end{equation}.
\item Provides the \texttt{jam\_axi\_proj} function with the stellar light MGE, trial total density MGE, trail BH mass, the reconstructed PSF obtained from of the DAP files, $RMS_{LOS}$, $RMS_{LOS}$ error, the rotated coordinates of the velocity map, a trail inclination to generate a spherical JAM model of $RMS_{LOS}$.
\item Returns the $\chi^2$ fit of the JAM model to the $RMS_{LOS}$ map.
\end{itemize}

A best guess, uncertainties, and bounds for each trial parameter is required by \texttt{AdaMet} for the optimization. The best guess of $M_*/L_*$ is the same as described in Section~\ref{fireflyer}. The relative uncertainty in the $M_*/L_*$ is then determined by adding in quadrature the relative uncertainty in the FIREFLY stellar mass bins with the relative uncertainty in the measured flux of the target. The bounds of $M_*/L_*$ are set within three times the uncertainty. The uncertainty in the stellar mass is then determined from the uncertainty in $M_*/L_*$. This range in the $M_*/L_*$ boundary ensures that the best fit will be achieved regardless of whether the $M_*/L_*$ was computed using slightly different cosmological parameters or a different modelling method of the total luminosity.

SILO can use equation 8 from~\cite{2009ApJ...704.1135B} to project the BH mass given a projection of the total mass of the galaxy. The total mass of the galaxy can be obtained by adding the measurement of the total stellar mass with the projection of the total physical mass of the DM halo. The total physical mass of the DM halo is projected by multiplying the total stellar mass from FIREFLY with the DM halo mass to stellar M/L predicted using equation 6 and Table 2 from~\cite{2020A&A...634A.135G}. The uncertainties in the total physical mass of the DM halo are then determined using the uncertainties stated in Table 2 from~\cite{2020A&A...634A.135G} with the uncertainties in the stellar mass. The galaxy's total mass is then projected as the sum of the stellar mass from FIREFLY with the projection of the total physical mass of the DM halo. The uncertainties in the galaxy's total mass are projected using the uncertainties in both the stellar mass and the total physical mass of the DM halo. SILO then computes the best guess of the trial BH mass using the projection of the galaxy's total mass in equation 8 of~\cite{2009ApJ...704.1135B}. The uncertainty in the trial BH mass is determined from the uncertainty in the galaxy's total mass with the uncertainty stated in equation 8 of~\cite{2009ApJ...704.1135B}. The bounds of the trial BH mass are determined from three times the uncertainty in the galaxy's total mass with three times the uncertainty stated in equation 8 of~\cite{2009ApJ...704.1135B}. While \cite{2020A&A...634A.135G} assume Planck cosmological constraints~\citep{2016A&A...594A..13P} and \cite{2009ApJ...704.1135B} assumes $H_0 = 70\ \mathrm{km\,s}^{-1}\mathrm{Mpc}^{-1}$ and $\Omega_m = 0.3$, these differences from WMAP cosmological constraints introduce an insignificant bias relative to the parameter boundaries set for the fitting process.

The best guess of the radial anisotropy is set as isotropic with uncertainty and bounds set as the full range of the possible anisotropy. The best guess of the perturbation to the PA angle is set as zero. The uncertainty and bounds in the perturbation to the PA angle are set as one and three times the uncertainty in the PA measurement from \texttt{fit\_kinematic\_pa}, respectively.

The optimizer then performs 10,000 iterations to fit the trial $q$, $M_*/L_*$, black hole mass, radial anisotropy, and the PA's perturbation using the trail function. The upper limit of the total density MGE is computed using the upper uncertainty in the fitted parameters in the trial function. The map of the upper limit of the mass enclosed within a test radius can then be projected using equation 11 from~\cite{2013MNRAS.432.1709C} with the upper limit of the total density MGE as input. These first measurements yields approximations of the probable strong lensing regime for the inspector to use in determining a radial cut of eight arcseconds is sufficient to consider candidate background galaxies beyond this radius as highly unlikely to be lensed. Finally, the optimizer process is repeated using a computationally-expensive 100,000 iterations for targets with candidate background galaxies located within eight arcseconds, which improved computation better evaluates the probable strong lensing regime.

\end{document}